\documentclass[10pt]{article}
\usepackage{helvet, graphicx}
\usepackage{cite}
\usepackage[hidelinks]{hyperref}
\usepackage[usenames]{color}
\usepackage{amsmath,amssymb,amsfonts, braket}
\usepackage{bm}
\usepackage[margin=0.75in]{geometry}
\usepackage[T1]{fontenc}
\usepackage[font=small]{caption}
\usepackage{graphicx} 

\usepackage{bibspacing}
\usepackage{setspace}

\title{Cavity-coupled telecom atomic source in silicon }
\author{Adam Johnston$^{1,2*}$, Ulises Felix-Rendon$^{1,2*}$, Yu-En Wong$^{1,2}$\thanks{These authors contributed equally to this work}~, Songtao Chen$^{1,3}$\thanks{songtao.chen@rice.edu}}

\date{%
    \centering
    \textit{\normalsize $^1$Department of Electrical and Computer Engineering, Rice University, Houston, TX 77005, USA}\\
    \textit{\normalsize $^2$Applied Physics Graduate Program, Smalley-Curl Institute, Rice University, Houston, TX 77005, USA}\\
    \textit{\normalsize $^3$Smalley-Curl Institute, Rice University, Houston, TX 77005, USA}\\[2ex]%
    \vspace{-\baselineskip}
    \vspace{-\baselineskip}
}

\begin{document}

\maketitle

\begin{abstract}
Atomic defects in solid-state materials are promising candidates for quantum interconnect and networking applications. Recently, a series of atomic defects have been identified in the silicon platform, where scalable device integration can be enabled by mature silicon photonics and electronics technologies.
In particular, T centers hold great promise due to their telecom band optical transitions and the doublet ground state electronic spin manifold with long coherence times. However, an open challenge for advancing the T center platform is to enhance its weak and slow zero phonon line emission. In this work, we demonstrate the cavity-enhanced fluorescence emission from a single T center. This is realized by integrating single T centers with a low-loss, small mode-volume silicon photonic crystal cavity, which results in an enhancement of the fluorescence decay rate by a factor of $F = 6.89$. Efficient photon extraction enables the system to achieve an average photon outcoupling rate of 73.3 kHz at the zero phonon line. The dynamics of the coupled system is well modeled by solving the Lindblad master equation.
These results represent a significant step towards building efficient T center spin-photon interfaces
for quantum information processing and networking applications. 

\end{abstract}

\vspace{10pt}

Optically interfaced atomic defects in solid-state materials are promising candidates for a variety of quantum technologies \cite{awschalom2018quantum}.
In particular, nitrogen and silicon vacancy centers in diamond have been used to demonstrate milestone results for fiber-based quantum networks, including spin-photon entanglement \cite{togan2010quantum}, deterministic entanglement generation between remote spins \cite{bernien2013heralded}, quantum state teleportation \cite{pfaff2014unconditional}, and memory-enhanced communication \cite{bhaskar2020experimental}. However, these defects have optical transitions at the visible or near-infrared spectral range resulting in large fiber transmission loss, requiring nonlinear frequency conversion \cite{li2016efficient} to extend the network range.
Significant progress has been made towards utilizing atomic defects with telecom optical transitions, leading to the discovery of single vanadium ions in silicon carbide \cite{wolfowicz2020vanadium} and single erbium ions in yttrium orthosilicate \cite{dibos2018atomic}, as well as recent demonstration of indistinguishable telecom photon generation from single erbium ions \cite{ourari2023indistinguishable}.

Early exploration into scalable fabrication of atomic defects with photonic structures relies on heterogeneous material integration and typically involves pick-and-place type of fabrication procedures \cite{kim2017hybrid, dibos2018atomic, wan2020large}. On the other hand, silicon-on-insulator (SOI) is a mature and scalable platform to enable large-scale monolithic photonic and electronic device integration.
Telecom-interfaced solid-state spins in silicon can thus benefit from the technological advantages of the SOI platform, and be utilized for realizing large-scale spin-based integrated quantum photonic chips  \cite{pelucchi2022potential}.
Moreover, silicon can be isotopically enriched to create a ``semiconductor vacuum'' \cite{steger2012quantum} for lowering the magnetic noise generated from the $^{29}$Si nuclear spin bath.

Beyond efforts towards optical addressing of erbium ions in silicon \cite{yin2013optical, gritsch2022narrow, berkman2023millisecond}, multiple novel atomic defect centers in silicon, including $C$, $G$, $T$, and $W$ centers \cite{chartrand2018highly, beaufils2018optical, redjem2020single, komza2022indistinguishable, prabhu2023individually, bergeron2020silicon, higginbottom2022optical, deabreu2023waveguide, tait2020microring, baron2022detection},
have been experimentally identified recently towards quantum information applications. Among them, T centers are particularly promising due to their telecom O-band optical transitions, doublet ground state spin manifold, and long spin coherence times in an enriched $^{28}$Si sample \cite{bergeron2020silicon}. Single T centers have been isolated in micropuck \cite{higginbottom2022optical} and waveguide \cite{lee2023high} structures. To further advance the single T center platform for quantum networking applications,
challenges remain to enhance its weak and slow coherent emission at the zero phonon line (ZPL). The cavity-induced Purcell effect \cite{purcell1995spontaneous} has been widely used for enhancing the fluorescence emission of various atomic defects in solids, including $G$ centers \cite{lefaucher2023cavity, redjem2023all, saggio2023cavity}. 

In this work, we demonstrate Purcell enhancement of a single T center in a low-loss, small mode-volume silicon photonic crystal (PC) cavity. When the cavity is tuned into resonance with the single T center we observe an enhancement of its fluorescence decay rate by a factor of $F = 6.89$, shortening the single T center's lifetime to 136.4 $\pm$ 0.6 ns. Leveraging the nanophotonic circuit and an angle-polished fiber for light coupling \cite{chen2021hybrid}, the system detection efficiency reaches $\eta_\text{sys}$ = 9.1\%, representing the probability of a single photon emitted into the cavity being registered by the detector. This efficiency is 20-fold larger than that achieved in a typical confocal-type measurement system for T centers. We probe single T centers in the device using time-resolved photoluminescence excitation (PLE) spectroscopy. Under the pulsed excitation, the system can detect 0.01 ZPL photon per excitation, reaching an average photon count rate of 73.3 kHz, which is
about two orders of magnitude improvement from previously reported emission rate for single T centers in the waveguide \cite{lee2023high}. By solving the Lindblad master equation, we develop a numerical model to describe the coupling dynamics between the single T center and the cavity, and extract cavity-QED parameters as well as T center pure dephasing rate ($\Gamma_\text{d}$) and spectral diffusion ($\Gamma_\text{sd}$).
This work represents a key step towards utilizing single T centers in silicon for quantum information applications.

\begin{figure}[ht]
	\centering
    \includegraphics{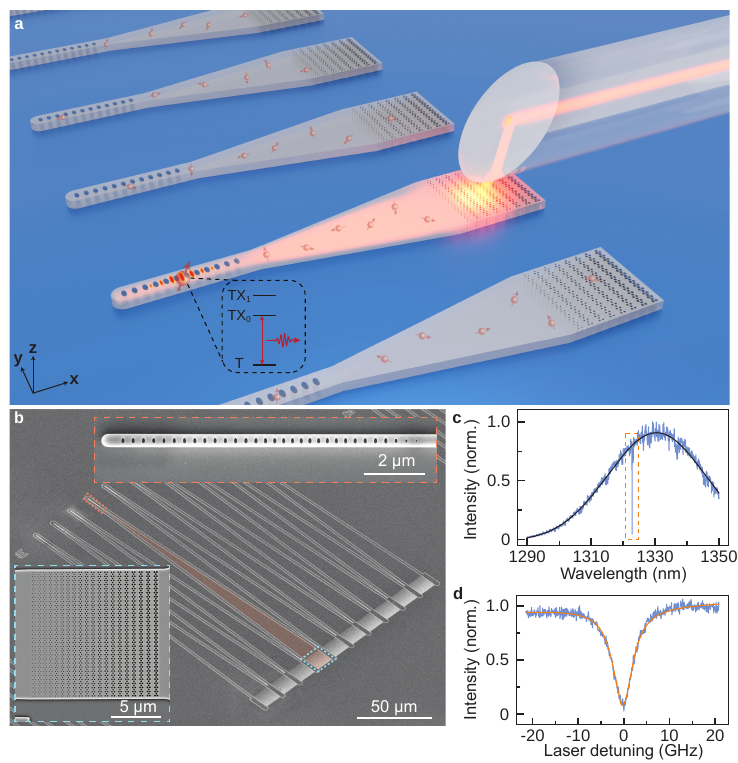}
    \caption{
    \textbf{Efficient optical coupling to single T centers in a silicon photonic cavity.} 
    \textbf{a.} Schematic illustration of fiber coupling to the nanophotonic circuits. Each PC cavitiy and subwavelength GC (with a width of 13.1 $\mu$m) are connected via a linearly tapered waveguide (200 $\mu$m in length). The GC allows efficient light coupling with an angle-polished fiber.
    ($x$, $y$, $z$) refer to ($[0\bar{1}0]$, $[100]$, $[001]$) directions of the silicon crystal. T centers (solid red balls with arrows) are uniformly generated across the whole SOI device layer using the ion implantation method. Inside the dashed square, the simplified electronic level structure of the single T center is shown \cite{bergeron2020silicon}. 
    \textbf{b.} Scanning electron microscope image of a block of 10 devices. An individual device (orange shaded) consists of a PC cavity on the upper left end (inside the orange dashed rectangle) and a GC on the lower right end (inside the blue dashed rectangle).
    \textbf{c.} Measured GC coupling spectrum (blue), with a Gaussian (black) fitted FWHM linewidth of 36.1 nm.
    The orange dashed box encloses a PC cavity.
    \textbf{d.} Reflection spectrum of the cavity (blue) shown in panel \textbf{c}, with a Lorentzian (orange) fitted FWHM linewidth of $5.22 \pm 0.04$ GHz. This cavity is used in later experiments. Results shown in panel \textbf{c} and \textbf{d} are measured at $T = 3.4$ K.
    }
    \label{fig:Figure1}
\end{figure}

\begin{figure}[ht]
	\centering
    \includegraphics{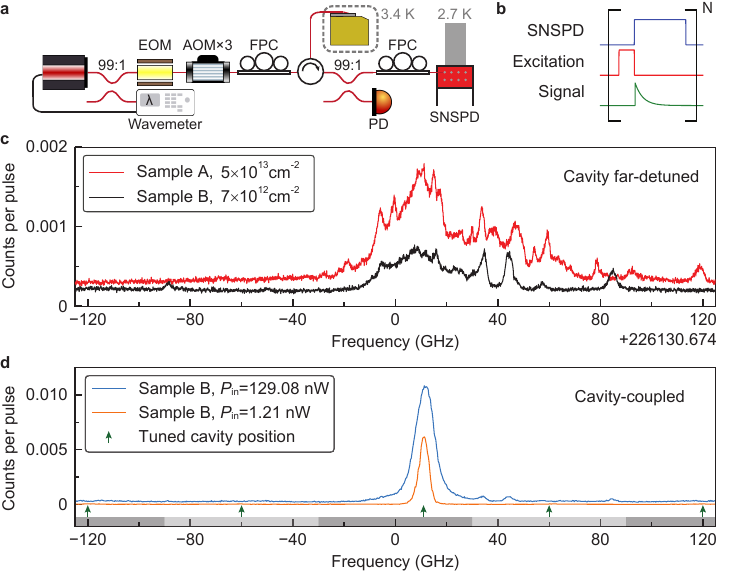}
    \caption{
    \textbf{Photoluminescence excitation spectroscopy for T centers.}
    \textbf{a.} Simplified schematic of the experimental setup for performing PLE spectroscopy. 
    Short excitation pulses
    are generated via a series of acousto-optic modulators (AOMs) and an electro-optic modulator (EOM) from a stabilized laser. The EOM is also used to generate laser sidebands. Fluorescence photons are redirected via an optical circulator to a SNSPD in another cryostat with $T = 2.7$ K for detection. A small portion of the signal enters a photodiode (PD) for monitoring cavity reflection. FPC: fiber polarization controller. 
    \textbf{b.} PLE pulse sequence. The SNSPD is gated to prevent detector latching due to the strong laser excitation pulses.
    \textbf{c.} PLE spectrum for two different implanted SOI samples under similar excitation conditions when the cavity is far-detuned ($\Delta_{ac} > 200$ GHz). The vertical axis indicates the average number of photons detected in an 5.55 $\mu$s integration window after each excitation pulse. Both samples are implanted with 1:1 ratio of $^{12}$C and $^1$H, and the implantation fluences are labeled in the legend. All the frequencies in this paper, if not specified, are offset from a reference $f_0 = 226130.674$ GHz (1325.749 nm or 935.201 meV).
    \textbf{d.} PLE spectrum for sample B at different excitation powers when the cavity is tuned within the scanning range. Each tuned cavity position is indicated by a green arrow, with the corresponding laser scanning range shaded on the $x$-axis under the arrow. The spectrum shows a dominant peak, which we interpret as a single T center coupled to the cavity.
    }
	\label{fig:Figure2}
\end{figure}

Our experimental configuration is outlined in Fig.~\ref{fig:Figure1}a. The nanophotonic devices are fabricated on a SOI sample which is situated on the cold finger of a closed-cycle cryostat ($T = 3.4$ K). Each device consists of a subwavelength grating coupler (GC) \cite{liu2010high} and a one-dimensional PC cavity, which are connected by a linearly tapered waveguide (Fig.~\ref{fig:Figure1}b). Optical coupling to the PC cavities is accomplished by using an angle-polished fiber via the GC with an one-way coupling efficiency of $\eta_\text{GC} = 46.1$\% at 1326 nm (Fig.~\ref{fig:Figure1}c). The fiber is mounted on a three-axis translation stage for optimizing the coupling. The PC cavity used in this work (Fig.~\ref{fig:Figure1}d) has a quality factor $Q = 4.3 \times 10^4$. Fluorescence from the T center is detected by a fiber-coupled superconducting nanowire single photon detector (SNSPD) located in a separate cryostat (Fig.~\ref{fig:Figure2}a). To match the atomic transition, we coarsely red-tune the cavity resonance by condensing nitrogen gas onto the surface of the device; we fine blue-tune the cavity resonance by sending laser pulses with a high optical power into the cavity \cite{suppinfo}. 

 
T centers are generated in the middle of the device layer (220 nm in thickness) of the SOI samples using the ion implantation method \cite{macquarrie2021generating}. Each SOI sample is implanted sequentially with 1:1 ratio of 35 keV $^{12}$C and 8 keV $^1$H ions, with rapid thermal annealing performed in-between and after the implantation steps \cite{suppinfo}. The two samples shown in this work have an implantation fluence of $5 \times 10^{13}$ cm$^{-2}$ (sample A) and $7 \times 10^{12}$ cm$^{-2}$ (sample B), respectively. To search for T centers, we perform time-resolved PLE spectroscopy by scanning the wavemeter-stabilized laser around the reported inhomogeneous center of T centers in silicon \cite{bergeron2020silicon,macquarrie2021generating, higginbottom2022optical}, with the pulse sequence shown in Fig.~\ref{fig:Figure2}b.

First, we measure the spectrum with the cavity far-detuned from the scan range (Fig.~\ref{fig:Figure2}c). The inhomogeneous distribution linewidth is about $\Gamma_\text{inh} \approx  29$ GHz. 
Isolated peaks can be observed away from the inhomogeneous center, which we interpret as the optical transitions of T centers that are likely located inside the taper waveguide. These peaks have an average full width half maximum (FWHM) linewidth of $2.40 \pm 0.86$ GHz, and a fluorescence lifetime of $836.8 \pm 57.3$ ns \cite{suppinfo}, which is slightly shorter than the bulk T centers' lifetime of 940 ns \cite{bergeron2020silicon}. The estimated T center densities are $\sim 1 \times 10^{12}$ cm$^{-3}$ and $\sim 3 \times 10^{11}$ cm$^{-3}$ for sample A and B, respectively \cite{suppinfo}.
For a specific waveguide-coupled T center (at 46 GHz) in sample B, we use the bounded T center quantum efficiency (discussed below) to estimate its emission efficiency to the waveguide mode as $2.6 \% \leq \eta_\text{wg} \leq 10.9$\% \cite{suppinfo}.

\begin{figure}[ht]
	\centering
    \includegraphics{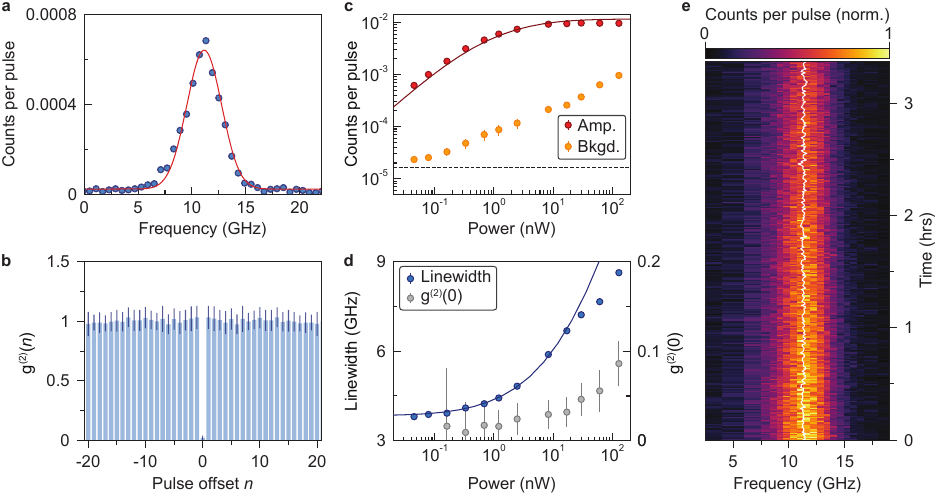}
    \caption{\textbf{Cavity-coupled single T center characterizations.} 
    \textbf{a.} PLE spectrum of the cavity-coupled single T center under an excitation power $P_\text{in}$ = 0.04 nW. 
    The red line shows the Gaussian fitting.
    \textbf{b.} Second-order autocorrelation measurement for the same T center ($P_\text{in}$ = 2.47 nW) shows a strong antibunching.
    All the photons detected after each excitation pulse are binned into a single time bin. The horizontal axis shows the autocorrelation offset in unit of the pulse repetition period (8 $\mu$s).
    \textbf{c.} Gaussian-fitted amplitude and background of single T-center PLE spectrum at different $P_\text{in}$. The background has the main contribution from other weakly-coupled T centers beyond the SNSPD dark counts (3 Hz, black dashed line). The saturation behavior is well described by the numerical model (red line).
    \textbf{d.} Spectrum linewidth of the single T center and the 
    $g^2(0)$ at different $P_\text{in}$. The blue line shows the numerical calculation results.
    For results shown in panel \textbf{a} $-$ \textbf{d}, the cavity is tuned into resonance with the T center transition ($\Delta_{ac} = 0$).
    \textbf{e.} Spectral diffusion of the single T center ($P_\text{in}$ = 8.31 nW). White line shows the center of the spectrum at each iteration (with a duration of 42 seconds) of the experiment.
    }
	\label{fig:Figure3}
\end{figure}

Next, we scan the laser frequency with the cavity tuned in-range to obtain the cavity-coupled PLE spectrum (Fig.~\ref{fig:Figure2}d). In sample B, a new T center peak at the inhomogeneous center emerges with its fluorescence significantly surpassing all other peaks.
This cavity-coupled T center has a FWHM linewidth of 
$\Gamma = 3.81 \pm 0.07$ GHz under a low excitation power
(Fig.~\ref{fig:Figure3}a). To verify the peak originates from a single T center, we measure the second-order autocorrelation function $g^{(2)}$ using all the detected fluorescence photons after each excitation pulse (Fig.~\ref{fig:Figure3}b). Photon antibunching is observed with the value $g^{(2)}(0) = 0.024 \pm 0.018$, which
confirms the majority of the detected photons come from a single emitter. Autocorrelation measurements for the single T center can show bunching ($g^{(2)}(n)>1$, when $|n|\geq1$) under certain excitation conditions, which we speculate to be caused by the spectral diffusion \cite{suppinfo}.

The emission amplitude of the cavity-coupled single T center saturates at 0.01 photons per excitation pulse (Fig.~\ref{fig:Figure3}c). Both the saturation and power dependent linewidth (Fig.~\ref{fig:Figure3}d) are well described by the numerical modeling (discussed below). The measured $g^{(2)}(0)$ at higher powers (Fig.~\ref{fig:Figure3}d) is limited by the accidental coincidences from background T centers' emission \cite{suppinfo}.
To characterize the spectral diffusion, we monitor the spectrum of the cavity-coupled single T center over a few hours time span by taking repetitive PLE scans (Fig.~\ref{fig:Figure3}e), which reveals a spectrum-center distribution of $11.30 \pm 0.13$ GHz. 
A similar level of spectral diffusion is observed for waveguide-coupled T centers \cite{suppinfo}. We note that this method only provides a lower bound of the $\Gamma_\text{sd}$ due to the limited experiment repetition rate. We later turn to the numerical modeling to extract $\Gamma_\text{sd}$.

We apply a magnetic field ($B$) up to 300 mT along silicon $[100]$ direction aiming to split the single T center line. We note that we have not been able to observe unambiguous Zeeman splitting using simultaneous two-tone laser sideband excitation \cite{suppinfo}, which is likely due to the limited splitting compared with the broad single T center linewidth.
When using the single-tone laser excitation, the PLE amplitude decreases at increasing $B$ field due to spin polarization \cite{macquarrie2021generating}. We model this behavior \cite{suppinfo} to extract the difference of the excited- and ground-state $g$-factors $|\Delta_g| = 0.55 \pm 0.04$, which matches with one of the two predicted $|\Delta_g|$ values for T centers under a $B$ field along the silicon $[100]$ direction \cite{macquarrie2021generating}.



\begin{figure}[ht]
	\centering
    \includegraphics{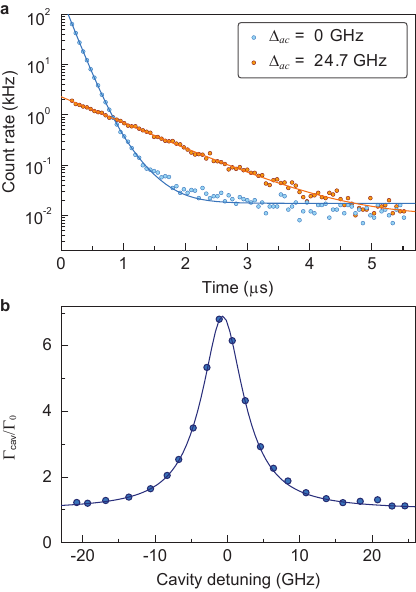}
    \caption{\textbf{Cavity-enhanced fluorescence emission of the single T center.} \textbf{a.} Time-resolved fluorescence for the cavity-coupled single T center under saturation ($P_\text{in} = $ 17.01 nW) 
    with different cavity detunings. The exponential fitting (blue and orange lines) reveals decay lifetimes of 136.4 $\pm$ 0.6 ns and 835.2 $\pm$ 3.1 ns for $\Delta_{ac} = 0$ GHz and $\Delta_{ac} = 24.7$ GHz, respectively.
    \textbf{b.} Decay rate enhancement at different cavity detunings with the laser fixed at the T center transition under an excitation power $P_\text{in} = 1.21$ nW. To gain better tuning accuracy, the cavity resonance is first red-tuned to $\Delta_{ac} = -20.8$ GHz, and subsequently blue-tuned.
    The blue line shows the numerical calculation results based on solving the Lindblad master equation.
    }
	\label{fig:Figure4}
\end{figure}

Lastly, we study the cavity-QED of the coupled system. When the cavity is tuned into resonance, the single T center's fluorescence lifetime is shortened to $136.4 \pm 0.6$ ns (Fig.~\ref{fig:Figure4}a), which is $6.89 \pm 0.03$ times faster than the bulk lifetime of $1/\Gamma_0 = 940$ ns \cite{bergeron2020silicon}. Leveraging this enhanced decay, we extract an average ZPL photon outcoupling rate of 73.3 kHz at saturation for the cavity-coupled single T center.
To confirm the enhancement originates from the cavity coupling, we measure the fluorescence decay rate $\Gamma_\text{cav}$ at different cavity detunings ($\Delta_{ac}$) (Fig.~\ref{fig:Figure4}b), which can be described as $\Gamma_\text{cav}/\Gamma_0 = P_t/[1+(2\Delta_{ac}/\tilde{\kappa})^2] + \Gamma_{\infty}/\Gamma_0$, where $P_t = 5.88 \pm 0.04$ is the Purcell factor describing the fluorescence decay enhancement due to the cavity, $\Gamma_{\infty} = (1.03 \pm 0.02) \Gamma_0 \approx \Gamma_0$ is the asymptotic decay rate at large detunings, and $\tilde{\kappa}/2\pi = 7.11 \pm 0.09$ GHz is the characteristic linewidth. 
To explain the deviation of $\tilde{\kappa}$ from the cavity linewidth $\kappa/2\pi = 5.22$ GHz, we turn to numerical calculations by solving the Lindblad master equation. Beyond the cavity and atomic loss channels, we also incorporate the pure dephasing and spectral diffusion processes \cite{suppinfo}. The dynamics of the coupled system can be described by the Jaynes-Cummings Hamiltonian of the form,

\begin{equation}
    H/\hbar=\Delta_a\sigma_+\sigma_- 
    + \Delta_c a^\dagger a
    + g(\sigma_+ a + \sigma_-a^\dagger)
    + \frac{\Omega}{2}(\sigma_+ + \sigma_-),
    \label{eq:RWA}
\end{equation}

\noindent which assumes rotating wave approximation in the rotating frame of the laser field ($\omega_L$). Here $\Delta_a = \omega_a - \omega_L$ and $\Delta_c = \omega_c - \omega_L$ are, respectively, the detunings of the laser from the T center transition $\omega_a$ and the cavity resonance $\omega_c$; $g$ is the coupling rate between the single T center and the cavity mode, and $\Omega$ is the optical Rabi frequency. The global fitting of the experimental data based on the numerical calculations \cite{suppinfo} reveals the cavity-QED parameters $(g, \kappa, \Gamma_0) = 2\pi \times$(42.4 MHz, 5.22 GHz, 169.3 kHz), an excited-state dephasing rate $2\Gamma_\text{d} = 2\pi\times1.29$ GHz and a spectral diffusion $\Gamma_\text{sd} = 2\pi\times 1.69$ GHz. The characteristic linewidth $\tilde{\kappa}$ has contributions from the cavity linewidth $\kappa$ as well as $\Gamma_\text{d}$ and $\Gamma_\text{sd}$.
The model can simultaneously capture the saturation (Fig.~\ref{fig:Figure3}c), power-dependent linewidth (Fig.~\ref{fig:Figure3}d), and detuning-dependent fluorescence decay (Fig.~\ref{fig:Figure4}b) results. Beyond the dephasing and the spectral diffusion, the single T center linewidth also has a minor contribution from the thermal broadening, which we estimate as $\Gamma_\text{th}\sim 0.1$~GHz from the temperature-dependent linewidth measurements \cite{suppinfo}.

To find out the Purcell enhancement of the single T center's ZPL ($P_\text{ZPL}$) and its quantum efficiency ($\eta_\text{QE}$), we express the total emission rate of a single T center in absence of a cavity as the summation of the emission rates into the ZPL and the phonon sideband (PSB), as well as nonradiative relaxation \cite{johnson2015tunable}, $\Gamma_0 = \gamma_\text{ZPL} + \gamma_\text{PSB} + \gamma_\text{nr}$. The Debye-Waller (DW) factor and the quantum efficiency can then be defined as DW = $\gamma_\text{ZPL}/(\gamma_\text{ZPL} + \gamma_\text{PSB})$ and $\eta_\text{QE} = (\gamma_\text{ZPL} + \gamma_\text{PSB})/\Gamma_0$, respectively. When the cavity is tuned into resonance with the T center ZPL, the cavity enhanced decay rate is $\Gamma_\text{cav} = (P_\text{ZPL}+1)\gamma_\text{ZPL} + \gamma_\text{PSB} + \gamma_\text{nr}$, where $P_\text{ZPL} = P_t/(\text{DW}\eta_\text{QE})$ is the Purcell factor describing the enhancement of the ZPL. We can thus put a lower bound on the $P_\text{ZPL} \geq P_t/\text{DW} = 25.6$, using the reported DW of 23\% \cite{bergeron2020silicon}. For simplicity, we neglect the potential suppression of the $\gamma_\text{PSB}$ due to the cavity \cite{johnson2015tunable}. The ratio of the single T center ZPL emission coupled to the resonant cavity mode can be estimated as $\beta = P_\text{ZPL}\gamma_\text{ZPL}/\left[(P_\text{ZPL}+1)\gamma_\text{ZPL} + \gamma_\text{PSB}+\gamma_\text{nr}\right] = P_t/(P_t+1) = 85.5$\%. Due to sub-optimal positioning of the single T center inside the cavity and imperfect dipole alignment with the local cavity electrical field polarization, the $P_\text{ZPL}$ extracted from measurements should be smaller than the simulated Purcell ($P_\text{ZPL} \leq P_\text{ZPL}^\text{sim}$). This enables us to put a lower bound on the quantum efficiency $\eta_\text{QE} \geq  23.4$\% for the single T center \cite{suppinfo}.
We now discuss pathways to improve the performance of the cavity-coupled single T center system. Due to technical difficulties, the first 170 ns of the fluorescence signal cannot be collected. The fluorescence count level can be improved by a factor of 2.5 by using AOMs with a faster switching speed. We note that the linewidth of the observed single T center is significantly larger than the Purcell enhanced lifetime-limited linewidth (2$\pi\times1.2$ MHz). One culprit is the fast dephasing process, which necessitates further investigation to reveal its origin. 
Significant reduction of $\Gamma_\text{inh}$ down to 33 MHz has been demonstrated for ensemble T centers in enriched $^{28}$Si \cite{bergeron2020silicon}. In future work, SOI samples with an enriched silicon device layer can be prepared via molecular beam epitaxy \cite{liu202228silicon} to minimize the dephasing. Furthermore, local electrodes can be fabricated on the SOI device layer for implementing electrical field control to minimize the spectral diffusion via \emph{in situ} tuning \cite{acosta2012dynamic} or depletion of the charge noises \cite{anderson2019electrical}. Lastly, focused-ion-beam-based \cite{schroder2017scalable} and masked \cite{toyli2010chip} ion implantation can be leveraged to increase the yield of T center generation at the cavity center. 



In summary, we have demonstrated enhanced light-matter interaction for a single T center by integrating it with a silicon nanophotonic cavity.
This work opens the door to utilize single T centers in silicon for quantum information processing and networking applications. With realistic improvements in the quality factor of the optical cavity ($Q = 5\times10^5$) and narrower linewidth in an enriched sample ($\Gamma$ $\sim$ 10 MHz), a large atom-cavity cooperativity $C \geq 29$ can be expected, which can lead to applications for high-fidelity dispersive spin readout \cite{nguyen2019integrated} and cavity-mediated spin-spin interactions \cite{evans2018photon}. Moreover, the current approach can enable parallel control and readout of multiple T centers in the cavity via the frequency domain addressing technique \cite{chen2020parallel}. Finally, leveraging the mature silicon photonics technology, small-footprint and scalable T-center-spin-based silicon quantum photonic chips \cite{pelucchi2022potential} may be envisioned. 



\textit{Note}: While finalizing this manuscript, we became aware of a related publication on detection of a single T center coupled to a cavity using above-band excitation \cite{islam2023cavity}.

\newpage
\textbf{\flushleft Acknowledgements}\\
{We gratefully acknowledge Han Pu, Alexey Belyanin, and Tanguy Terlier for helpful discussions, and John Bartholomew for feedback on a manuscript draft. Support for this research was provided by the National Science Foundation (NSF, CAREER Award No. 2238298), the Robert A. Welch Foundation (Grant No. C-2134) and the Rice Faculty Initiative Fund. We acknowledge the use of cleanroom facilities supported by the Shared Equipment Authority at Rice University.} 

\textbf{\flushleft Data Availability}\\
{The datasets generated and/or analyzed during the current study are available from the corresponding author on reasonable request.}

\textbf{\flushleft Competing interests}\\
{The authors declare no competing interests.}

\clearpage

\clearpage

\clearpage

\clearpage

\newpage
\setlength\bibitemsep{0pt}
\bibliography{SingleT_references}

\begin{thebibliography}{10}
\expandafter\ifx\csname url\endcsname\relax
  \def\url#1{\texttt{#1}}\fi
\expandafter\ifx\csname urlprefix\endcsname\relax\def\urlprefix{URL }\fi
\providecommand{\bibinfo}[2]{#2}
\providecommand{\eprint}[2][]{\url{#2}}

\bibitem{awschalom2018quantum}
\bibinfo{author}{Awschalom, D.~D.}, \bibinfo{author}{Hanson, R.},
  \bibinfo{author}{Wrachtrup, J.} \& \bibinfo{author}{Zhou, B.~B.}
\newblock \bibinfo{title}{Quantum technologies with optically interfaced
  solid-state spins}.
\newblock \emph{\bibinfo{journal}{Nat. Photon.}} \textbf{\bibinfo{volume}{12}},
  \bibinfo{pages}{516--527} (\bibinfo{year}{2018}).

\bibitem{togan2010quantum}
\bibinfo{author}{Togan, E.} \emph{et~al.}
\newblock \bibinfo{title}{Quantum entanglement between an optical photon and a
  solid-state spin qubit}.
\newblock \emph{\bibinfo{journal}{Nature}} \textbf{\bibinfo{volume}{466}},
  \bibinfo{pages}{730--734} (\bibinfo{year}{2010}).

\bibitem{bernien2013heralded}
\bibinfo{author}{Bernien, H.} \emph{et~al.}
\newblock \bibinfo{title}{Heralded entanglement between solid-state qubits
  separated by three metres}.
\newblock \emph{\bibinfo{journal}{Nature}} \textbf{\bibinfo{volume}{497}},
  \bibinfo{pages}{86--90} (\bibinfo{year}{2013}).

\bibitem{pfaff2014unconditional}
\bibinfo{author}{Pfaff, W.} \emph{et~al.}
\newblock \bibinfo{title}{Unconditional quantum teleportation between distant
  solid-state quantum bits}.
\newblock \emph{\bibinfo{journal}{Science}} \textbf{\bibinfo{volume}{345}},
  \bibinfo{pages}{532--535} (\bibinfo{year}{2014}).

\bibitem{bhaskar2020experimental}
\bibinfo{author}{Bhaskar, M.~K.} \emph{et~al.}
\newblock \bibinfo{title}{Experimental demonstration of memory-enhanced quantum
  communication}.
\newblock \emph{\bibinfo{journal}{Nature}} \textbf{\bibinfo{volume}{580}},
  \bibinfo{pages}{60--64} (\bibinfo{year}{2020}).

\bibitem{li2016efficient}
\bibinfo{author}{Li, Q.}, \bibinfo{author}{Davan{\c{c}}o, M.} \&
  \bibinfo{author}{Srinivasan, K.}
\newblock \bibinfo{title}{Efficient and low-noise single-photon-level frequency
  conversion interfaces using silicon nanophotonics}.
\newblock \emph{\bibinfo{journal}{Nat. Photon.}} \textbf{\bibinfo{volume}{10}},
  \bibinfo{pages}{406--414} (\bibinfo{year}{2016}).

\bibitem{wolfowicz2020vanadium}
\bibinfo{author}{Wolfowicz, G.} \emph{et~al.}
\newblock \bibinfo{title}{Vanadium spin qubits as telecom quantum emitters in
  silicon carbide}.
\newblock \emph{\bibinfo{journal}{Sci. Adv.}} \textbf{\bibinfo{volume}{6}},
  \bibinfo{pages}{eaaz1192} (\bibinfo{year}{2020}).

\bibitem{dibos2018atomic}
\bibinfo{author}{Dibos, A.}, \bibinfo{author}{Raha, M.},
  \bibinfo{author}{Phenicie, C.} \& \bibinfo{author}{Thompson, J.~D.}
\newblock \bibinfo{title}{Atomic source of single photons in the telecom band}.
\newblock \emph{\bibinfo{journal}{Phys. Rev. Lett.}}
  \textbf{\bibinfo{volume}{120}}, \bibinfo{pages}{243601}
  (\bibinfo{year}{2018}).

\bibitem{ourari2023indistinguishable}
\bibinfo{author}{Ourari, S.} \emph{et~al.}
\newblock \bibinfo{title}{Indistinguishable telecom band photons from a single
  {Er} ion in the solid state}.
\newblock \emph{\bibinfo{journal}{Nature}} \textbf{\bibinfo{volume}{620}},
  \bibinfo{pages}{977--981} (\bibinfo{year}{2023}).

\bibitem{kim2017hybrid}
\bibinfo{author}{Kim, J.-H.} \emph{et~al.}
\newblock \bibinfo{title}{Hybrid integration of solid-state quantum emitters on
  a silicon photonic chip}.
\newblock \emph{\bibinfo{journal}{Nano Lett.}} \textbf{\bibinfo{volume}{17}},
  \bibinfo{pages}{7394--7400} (\bibinfo{year}{2017}).

\bibitem{wan2020large}
\bibinfo{author}{Wan, N.~H.} \emph{et~al.}
\newblock \bibinfo{title}{Large-scale integration of artificial atoms in hybrid
  photonic circuits}.
\newblock \emph{\bibinfo{journal}{Nature}} \textbf{\bibinfo{volume}{583}},
  \bibinfo{pages}{226--231} (\bibinfo{year}{2020}).

\bibitem{pelucchi2022potential}
\bibinfo{author}{Pelucchi, E.} \emph{et~al.}
\newblock \bibinfo{title}{The potential and global outlook of integrated
  photonics for quantum technologies}.
\newblock \emph{\bibinfo{journal}{Nat. Rev. Phys.}}
  \textbf{\bibinfo{volume}{4}}, \bibinfo{pages}{194--208}
  (\bibinfo{year}{2022}).

\bibitem{steger2012quantum}
\bibinfo{author}{Steger, M.} \emph{et~al.}
\newblock \bibinfo{title}{Quantum information storage for over 180 s using
  donor spins in a $^{28}${Si} “semiconductor vacuum”}.
\newblock \emph{\bibinfo{journal}{Science}} \textbf{\bibinfo{volume}{336}},
  \bibinfo{pages}{1280--1283} (\bibinfo{year}{2012}).

\bibitem{yin2013optical}
\bibinfo{author}{Yin, C.} \emph{et~al.}
\newblock \bibinfo{title}{Optical addressing of an individual erbium ion in
  silicon}.
\newblock \emph{\bibinfo{journal}{Nature}} \textbf{\bibinfo{volume}{497}},
  \bibinfo{pages}{91--94} (\bibinfo{year}{2013}).

\bibitem{gritsch2022narrow}
\bibinfo{author}{Gritsch, A.}, \bibinfo{author}{Weiss, L.},
  \bibinfo{author}{Fr{\"u}h, J.}, \bibinfo{author}{Rinner, S.} \&
  \bibinfo{author}{Reiserer, A.}
\newblock \bibinfo{title}{Narrow optical transitions in erbium-implanted
  silicon waveguides}.
\newblock \emph{\bibinfo{journal}{Phys. Rev. X}} \textbf{\bibinfo{volume}{12}},
  \bibinfo{pages}{041009} (\bibinfo{year}{2022}).

\bibitem{berkman2023millisecond}
\bibinfo{author}{Berkman, I.~R.} \emph{et~al.}
\newblock \bibinfo{title}{Millisecond electron spin coherence time for erbium
  ions in silicon}.
\newblock \emph{\bibinfo{journal}{arXiv preprint arXiv:2307.10021}}
  (\bibinfo{year}{2023}).

\bibitem{chartrand2018highly}
\bibinfo{author}{Chartrand, C.} \emph{et~al.}
\newblock \bibinfo{title}{Highly enriched $^{28}${Si} reveals remarkable
  optical linewidths and fine structure for well-known damage centers}.
\newblock \emph{\bibinfo{journal}{Phys. Rev. B}} \textbf{\bibinfo{volume}{98}},
  \bibinfo{pages}{195201} (\bibinfo{year}{2018}).

\bibitem{beaufils2018optical}
\bibinfo{author}{Beaufils, C.} \emph{et~al.}
\newblock \bibinfo{title}{Optical properties of an ensemble of {G}-centers in
  silicon}.
\newblock \emph{\bibinfo{journal}{Phys. Rev. B}} \textbf{\bibinfo{volume}{97}},
  \bibinfo{pages}{035303} (\bibinfo{year}{2018}).

\bibitem{redjem2020single}
\bibinfo{author}{Redjem, W.} \emph{et~al.}
\newblock \bibinfo{title}{Single artificial atoms in silicon emitting at
  telecom wavelengths}.
\newblock \emph{\bibinfo{journal}{Nat. Electron.}}
  \textbf{\bibinfo{volume}{3}}, \bibinfo{pages}{738--743}
  (\bibinfo{year}{2020}).

\bibitem{komza2022indistinguishable}
\bibinfo{author}{Komza, L.} \emph{et~al.}
\newblock \bibinfo{title}{Indistinguishable photons from an artificial atom in
  silicon photonics}.
\newblock \emph{\bibinfo{journal}{arXiv preprint arXiv:2211.09305}}
  (\bibinfo{year}{2022}).

\bibitem{prabhu2023individually}
\bibinfo{author}{Prabhu, M.} \emph{et~al.}
\newblock \bibinfo{title}{Individually addressable and spectrally programmable
  artificial atoms in silicon photonics}.
\newblock \emph{\bibinfo{journal}{Nat. Commun.}} \textbf{\bibinfo{volume}{14}},
  \bibinfo{pages}{2380} (\bibinfo{year}{2023}).

\bibitem{bergeron2020silicon}
\bibinfo{author}{Bergeron, L.} \emph{et~al.}
\newblock \bibinfo{title}{Silicon-integrated telecommunications photon-spin
  interface}.
\newblock \emph{\bibinfo{journal}{PRX Quantum}} \textbf{\bibinfo{volume}{1}},
  \bibinfo{pages}{020301} (\bibinfo{year}{2020}).

\bibitem{higginbottom2022optical}
\bibinfo{author}{Higginbottom, D.~B.} \emph{et~al.}
\newblock \bibinfo{title}{Optical observation of single spins in silicon}.
\newblock \emph{\bibinfo{journal}{Nature}} \textbf{\bibinfo{volume}{607}},
  \bibinfo{pages}{266--270} (\bibinfo{year}{2022}).

\bibitem{deabreu2023waveguide}
\bibinfo{author}{DeAbreu, A.} \emph{et~al.}
\newblock \bibinfo{title}{Waveguide-integrated silicon {T} centres}.
\newblock \emph{\bibinfo{journal}{Opt. Express}} \textbf{\bibinfo{volume}{31}},
  \bibinfo{pages}{15045--15057} (\bibinfo{year}{2023}).

\bibitem{tait2020microring}
\bibinfo{author}{Tait, A.~N.} \emph{et~al.}
\newblock \bibinfo{title}{Microring resonator-coupled photoluminescence from
  silicon {W} centers}.
\newblock \emph{\bibinfo{journal}{J. Phys. Photonics}}
  \textbf{\bibinfo{volume}{2}}, \bibinfo{pages}{045001} (\bibinfo{year}{2020}).

\bibitem{baron2022detection}
\bibinfo{author}{Baron, Y.} \emph{et~al.}
\newblock \bibinfo{title}{Detection of single {W}-centers in silicon}.
\newblock \emph{\bibinfo{journal}{ACS Photon.}} \textbf{\bibinfo{volume}{9}},
  \bibinfo{pages}{2337--2345} (\bibinfo{year}{2022}).

\bibitem{lee2023high}
\bibinfo{author}{Lee, C.-M.} \emph{et~al.}
\newblock \bibinfo{title}{High-efficiency single photon emission from a silicon
  {T}-center in a nanobeam}.
\newblock \emph{\bibinfo{journal}{arXiv preprint arXiv:2308.04541}}
  (\bibinfo{year}{2023}).

\bibitem{purcell1995spontaneous}
\bibinfo{author}{Purcell, E.~M.}
\newblock \bibinfo{title}{Spontaneous emission probabilities at radio
  frequencies}.
\newblock In \emph{\bibinfo{booktitle}{Confined Electrons and Photons: New
  Physics and Applications}}, \bibinfo{pages}{839--839}
  (\bibinfo{publisher}{Springer}, \bibinfo{year}{1995}).

\bibitem{lefaucher2023cavity}
\bibinfo{author}{Lefaucher, B.} \emph{et~al.}
\newblock \bibinfo{title}{Cavity-enhanced zero-phonon emission from an ensemble
  of {G} centers in a silicon-on-insulator microring}.
\newblock \emph{\bibinfo{journal}{Appl. Phys. Lett.}}
  \textbf{\bibinfo{volume}{122}} (\bibinfo{year}{2023}).

\bibitem{redjem2023all}
\bibinfo{author}{Redjem, W.} \emph{et~al.}
\newblock \bibinfo{title}{All-silicon quantum light source by embedding an
  atomic emissive center in a nanophotonic cavity}.
\newblock \emph{\bibinfo{journal}{Nat. Commun.}} \textbf{\bibinfo{volume}{14}},
  \bibinfo{pages}{3321} (\bibinfo{year}{2023}).

\bibitem{saggio2023cavity}
\bibinfo{author}{Saggio, V.} \emph{et~al.}
\newblock \bibinfo{title}{Cavity-enhanced single artificial atoms in silicon}.
\newblock \emph{\bibinfo{journal}{arXiv preprint arXiv:2302.10230}}
  (\bibinfo{year}{2023}).

\bibitem{chen2021hybrid}
\bibinfo{author}{Chen, S.} \emph{et~al.}
\newblock \bibinfo{title}{Hybrid microwave-optical scanning probe for
  addressing solid-state spins in nanophotonic cavities}.
\newblock \emph{\bibinfo{journal}{Opt. Express}} \textbf{\bibinfo{volume}{29}},
  \bibinfo{pages}{4902--4911} (\bibinfo{year}{2021}).

\bibitem{liu2010high}
\bibinfo{author}{Liu, L.}, \bibinfo{author}{Pu, M.}, \bibinfo{author}{Yvind,
  K.} \& \bibinfo{author}{Hvam, J.~M.}
\newblock \bibinfo{title}{High-efficiency, large-bandwidth silicon-on-insulator
  grating coupler based on a fully-etched photonic crystal structure}.
\newblock \emph{\bibinfo{journal}{Appl. Phys. Lett.}}
  \textbf{\bibinfo{volume}{96}} (\bibinfo{year}{2010}).

\bibitem{suppinfo}
\bibinfo{title}{See the supplementary materials\!\!} .

\bibitem{macquarrie2021generating}
\bibinfo{author}{MacQuarrie, E.} \emph{et~al.}
\newblock \bibinfo{title}{Generating {T} centres in photonic
  silicon-on-insulator material by ion implantation}.
\newblock \emph{\bibinfo{journal}{N. J. of Phys.}}
  \textbf{\bibinfo{volume}{23}}, \bibinfo{pages}{103008}
  (\bibinfo{year}{2021}).

\bibitem{johnson2015tunable}
\bibinfo{author}{Johnson, S.} \emph{et~al.}
\newblock \bibinfo{title}{Tunable cavity coupling of the zero phonon line of a
  nitrogen-vacancy defect in diamond}.
\newblock \emph{\bibinfo{journal}{N. J. Phys.}} \textbf{\bibinfo{volume}{17}},
  \bibinfo{pages}{122003} (\bibinfo{year}{2015}).

\bibitem{liu202228silicon}
\bibinfo{author}{Liu, Y.} \emph{et~al.}
\newblock \bibinfo{title}{$^{28}${S}ilicon-on-insulator for optically
  interfaced quantum emitters}.
\newblock \emph{\bibinfo{journal}{J. Cryst. Growth}}
  \textbf{\bibinfo{volume}{593}}, \bibinfo{pages}{126733}
  (\bibinfo{year}{2022}).

\bibitem{acosta2012dynamic}
\bibinfo{author}{Acosta, V.} \emph{et~al.}
\newblock \bibinfo{title}{Dynamic stabilization of the optical resonances of
  single nitrogen-vacancy centers in diamond}.
\newblock \emph{\bibinfo{journal}{Phys. Rev. Lett.}}
  \textbf{\bibinfo{volume}{108}}, \bibinfo{pages}{206401}
  (\bibinfo{year}{2012}).

\bibitem{anderson2019electrical}
\bibinfo{author}{Anderson, C.~P.} \emph{et~al.}
\newblock \bibinfo{title}{Electrical and optical control of single spins
  integrated in scalable semiconductor devices}.
\newblock \emph{\bibinfo{journal}{Science}} \textbf{\bibinfo{volume}{366}},
  \bibinfo{pages}{1225--1230} (\bibinfo{year}{2019}).

\bibitem{schroder2017scalable}
\bibinfo{author}{Schr{\"o}der, T.} \emph{et~al.}
\newblock \bibinfo{title}{Scalable focused ion beam creation of nearly
  lifetime-limited single quantum emitters in diamond nanostructures}.
\newblock \emph{\bibinfo{journal}{Nat. Commun.}} \textbf{\bibinfo{volume}{8}},
  \bibinfo{pages}{15376} (\bibinfo{year}{2017}).

\bibitem{toyli2010chip}
\bibinfo{author}{Toyli, D.~M.}, \bibinfo{author}{Weis, C.~D.},
  \bibinfo{author}{Fuchs, G.~D.}, \bibinfo{author}{Schenkel, T.} \&
  \bibinfo{author}{Awschalom, D.~D.}
\newblock \bibinfo{title}{Chip-scale nanofabrication of single spins and spin
  arrays in diamond}.
\newblock \emph{\bibinfo{journal}{Nano Lett.}} \textbf{\bibinfo{volume}{10}},
  \bibinfo{pages}{3168--3172} (\bibinfo{year}{2010}).

\bibitem{nguyen2019integrated}
\bibinfo{author}{Nguyen, C.} \emph{et~al.}
\newblock \bibinfo{title}{An integrated nanophotonic quantum register based on
  silicon-vacancy spins in diamond}.
\newblock \emph{\bibinfo{journal}{Phys. Rev. B}}
  \textbf{\bibinfo{volume}{100}}, \bibinfo{pages}{165428}
  (\bibinfo{year}{2019}).

\bibitem{evans2018photon}
\bibinfo{author}{Evans, R.~E.} \emph{et~al.}
\newblock \bibinfo{title}{Photon-mediated interactions between quantum emitters
  in a diamond nanocavity}.
\newblock \emph{\bibinfo{journal}{Science}} \textbf{\bibinfo{volume}{362}},
  \bibinfo{pages}{662--665} (\bibinfo{year}{2018}).

\bibitem{chen2020parallel}
\bibinfo{author}{Chen, S.}, \bibinfo{author}{Raha, M.},
  \bibinfo{author}{Phenicie, C.~M.}, \bibinfo{author}{Ourari, S.} \&
  \bibinfo{author}{Thompson, J.~D.}
\newblock \bibinfo{title}{Parallel single-shot measurement and coherent control
  of solid-state spins below the diffraction limit}.
\newblock \emph{\bibinfo{journal}{Science}} \textbf{\bibinfo{volume}{370}},
  \bibinfo{pages}{592--595} (\bibinfo{year}{2020}).

\bibitem{islam2023cavity}
\bibinfo{author}{Islam, F.} \emph{et~al.}
\newblock \bibinfo{title}{Cavity enhanced emission from a silicon {T} center}.
\newblock \emph{\bibinfo{journal}{arXiv preprint arXiv:2310.13808}}
  (\bibinfo{year}{2023}).

\end{thebibliography}


\begin{thebibliography}{10}
\expandafter\ifx\csname url\endcsname\relax
  \def\url#1{\texttt{#1}}\fi
\expandafter\ifx\csname urlprefix\endcsname\relax\def\urlprefix{URL }\fi
\providecommand{\bibinfo}[2]{#2}
\providecommand{\eprint}[2][]{\url{#2}}

\bibitem{Mosor2005}
\bibinfo{author}{Mosor, S.} \emph{et~al.}
\newblock \bibinfo{title}{{Scanning a photonic crystal slab nanocavity by
  condensation of xenon}}.
\newblock \emph{\bibinfo{journal}{Appl. Phys. Lett.}}
  \textbf{\bibinfo{volume}{87}}, \bibinfo{pages}{141105}
  (\bibinfo{year}{2005}).

\bibitem{chan2009optical}
\bibinfo{author}{Chan, J.}, \bibinfo{author}{Eichenfield, M.},
  \bibinfo{author}{Camacho, R.} \& \bibinfo{author}{Painter, O.}
\newblock \bibinfo{title}{Optical and mechanical design of a “zipper”
  photonic crystal optomechanical cavity}.
\newblock \emph{\bibinfo{journal}{Opt. Express}} \textbf{\bibinfo{volume}{17}},
  \bibinfo{pages}{3802--3817} (\bibinfo{year}{2009}).

\bibitem{johnson2001block}
\bibinfo{author}{Johnson, S.~G.} \& \bibinfo{author}{Joannopoulos, J.~D.}
\newblock \bibinfo{title}{Block-iterative frequency-domain methods for
  {M}axwell’s equations in a planewave basis}.
\newblock \emph{\bibinfo{journal}{Opt. Express}} \textbf{\bibinfo{volume}{8}},
  \bibinfo{pages}{173--190} (\bibinfo{year}{2001}).

\bibitem{oskooi2010meep}
\bibinfo{author}{Oskooi, A.~F.} \emph{et~al.}
\newblock \bibinfo{title}{Meep: A flexible free-software package for
  electromagnetic simulations by the {FDTD} method}.
\newblock \emph{\bibinfo{journal}{Comput. Phys. Commun.}}
  \textbf{\bibinfo{volume}{181}}, \bibinfo{pages}{687--702}
  (\bibinfo{year}{2010}).

\bibitem{Bergeron2020PRXQuantum}
\bibinfo{author}{Bergeron, L.} \emph{et~al.}
\newblock \bibinfo{title}{Silicon-integrated telecommunications photon-spin
  interface}.
\newblock \emph{\bibinfo{journal}{PRX Quantum}} \textbf{\bibinfo{volume}{1}},
  \bibinfo{pages}{020301} (\bibinfo{year}{2020}).

\bibitem{dung2006local}
\bibinfo{author}{Dung, H.~T.}, \bibinfo{author}{Buhmann, S.~Y.} \&
  \bibinfo{author}{Welsch, D.-G.}
\newblock \bibinfo{title}{Local-field correction to the spontaneous decay rate
  of atoms embedded in bodies of finite size}.
\newblock \emph{\bibinfo{journal}{Phys. Rev. A}} \textbf{\bibinfo{volume}{74}},
  \bibinfo{pages}{023803} (\bibinfo{year}{2006}).

\bibitem{liu2010high}
\bibinfo{author}{Liu, L.}, \bibinfo{author}{Pu, M.}, \bibinfo{author}{Yvind,
  K.} \& \bibinfo{author}{Hvam, J.~M.}
\newblock \bibinfo{title}{High-efficiency, large-bandwidth silicon-on-insulator
  grating coupler based on a fully-etched photonic crystal structure}.
\newblock \emph{\bibinfo{journal}{Appl. Phys. Lett.}}
  \textbf{\bibinfo{volume}{96}} (\bibinfo{year}{2010}).

\bibitem{ding2013ultrahigh}
\bibinfo{author}{Ding, Y.}, \bibinfo{author}{Ou, H.} \&
  \bibinfo{author}{Peucheret, C.}
\newblock \bibinfo{title}{Ultrahigh-efficiency apodized grating coupler using
  fully etched photonic crystals}.
\newblock \emph{\bibinfo{journal}{Opt. Lett.}} \textbf{\bibinfo{volume}{38}},
  \bibinfo{pages}{2732--2734} (\bibinfo{year}{2013}).

\bibitem{halir2015waveguide}
\bibinfo{author}{Halir, R.} \emph{et~al.}
\newblock \bibinfo{title}{Waveguide sub-wavelength structures: a review of
  principles and applications}.
\newblock \emph{\bibinfo{journal}{Laser \& Photon. Rev.}}
  \textbf{\bibinfo{volume}{9}}, \bibinfo{pages}{25--49} (\bibinfo{year}{2015}).

\bibitem{halir2009waveguide}
\bibinfo{author}{Halir, R.} \emph{et~al.}
\newblock \bibinfo{title}{Waveguide grating coupler with subwavelength
  microstructures}.
\newblock \emph{\bibinfo{journal}{Opt. Lett.}} \textbf{\bibinfo{volume}{34}},
  \bibinfo{pages}{1408--1410} (\bibinfo{year}{2009}).

\bibitem{chen2021hybrid}
\bibinfo{author}{Chen, S.} \emph{et~al.}
\newblock \bibinfo{title}{Hybrid microwave-optical scanning probe for
  addressing solid-state spins in nanophotonic cavities}.
\newblock \emph{\bibinfo{journal}{Opt. Express}} \textbf{\bibinfo{volume}{29}},
  \bibinfo{pages}{4902--4911} (\bibinfo{year}{2021}).

\bibitem{MacQuarrie2021}
\bibinfo{author}{MacQuarrie, E.~R.} \emph{et~al.}
\newblock \bibinfo{title}{Generating {T} centres in photonic
  silicon-on-insulator material by ion implantation}.
\newblock \emph{\bibinfo{journal}{New J. Phys.}} \textbf{\bibinfo{volume}{23}},
  \bibinfo{pages}{103008} (\bibinfo{year}{2021}).

\bibitem{Ziegler2010}
\bibinfo{author}{Ziegler, J.~F.}, \bibinfo{author}{Ziegler, M.} \&
  \bibinfo{author}{Biersack, J.}
\newblock \bibinfo{title}{{SRIM} – {T}he stopping and range of ions in matter
  (2010)}.
\newblock \emph{\bibinfo{journal}{Nucl. Instrum. Methods Phys. Res. B: Beam
  Interact. Mater. At.}} \textbf{\bibinfo{volume}{268}},
  \bibinfo{pages}{1818--1823} (\bibinfo{year}{2010}).

\bibitem{Cho1985}
\bibinfo{author}{Cho, K.} \emph{et~al.}
\newblock \bibinfo{title}{Channeling effect for low energy ion implantation in
  {S}i}.
\newblock \emph{\bibinfo{journal}{Nucl. Instrum. Methods Phys. Res. B: Beam
  Interact. Mater. At.}} \textbf{\bibinfo{volume}{7-8}},
  \bibinfo{pages}{265--272} (\bibinfo{year}{1985}).

\bibitem{higginbottom2022optical}
\bibinfo{author}{Higginbottom, D.~B.} \emph{et~al.}
\newblock \bibinfo{title}{Optical observation of single spins in silicon}.
\newblock \emph{\bibinfo{journal}{Nature}} \textbf{\bibinfo{volume}{607}},
  \bibinfo{pages}{266--270} (\bibinfo{year}{2022}).

\bibitem{thurber1980resistivity}
\bibinfo{author}{Thurber, W.}, \bibinfo{author}{Mattis, R.},
  \bibinfo{author}{Liu, Y.} \& \bibinfo{author}{Filliben, J.}
\newblock \bibinfo{title}{Resistivity-dopant density relationship for
  boron-doped silicon}.
\newblock \emph{\bibinfo{journal}{J. Electrochem. Soc.}}
  \textbf{\bibinfo{volume}{127}}, \bibinfo{pages}{2291} (\bibinfo{year}{1980}).

\bibitem{uysal2023coherent}
\bibinfo{author}{Uysal, M.~T.} \emph{et~al.}
\newblock \bibinfo{title}{Coherent control of a nuclear spin via interactions
  with a rare-earth ion in the solid state}.
\newblock \emph{\bibinfo{journal}{PRX Quantum}} \textbf{\bibinfo{volume}{4}},
  \bibinfo{pages}{010323} (\bibinfo{year}{2023}).

\bibitem{dibos2018atomic}
\bibinfo{author}{Dibos, A.}, \bibinfo{author}{Raha, M.},
  \bibinfo{author}{Phenicie, C.} \& \bibinfo{author}{Thompson, J.~D.}
\newblock \bibinfo{title}{Atomic source of single photons in the telecom band}.
\newblock \emph{\bibinfo{journal}{Phys. Rev. Lett.}}
  \textbf{\bibinfo{volume}{120}}, \bibinfo{pages}{243601}
  (\bibinfo{year}{2018}).

\bibitem{sallen2010subnanosecond}
\bibinfo{author}{Sallen, G.} \emph{et~al.}
\newblock \bibinfo{title}{Subnanosecond spectral diffusion measurement using
  photon correlation}.
\newblock \emph{\bibinfo{journal}{Nat. Photon.}} \textbf{\bibinfo{volume}{4}},
  \bibinfo{pages}{696--699} (\bibinfo{year}{2010}).

\bibitem{redjem2020single}
\bibinfo{author}{Redjem, W.} \emph{et~al.}
\newblock \bibinfo{title}{Single artificial atoms in silicon emitting at
  telecom wavelengths}.
\newblock \emph{\bibinfo{journal}{Nat. Electron.}}
  \textbf{\bibinfo{volume}{3}}, \bibinfo{pages}{738--743}
  (\bibinfo{year}{2020}).

\bibitem{tiecke2014nanophotonic}
\bibinfo{author}{Tiecke, T.} \emph{et~al.}
\newblock \bibinfo{title}{Nanophotonic quantum phase switch with a single
  atom}.
\newblock \emph{\bibinfo{journal}{Nature}} \textbf{\bibinfo{volume}{508}},
  \bibinfo{pages}{241--244} (\bibinfo{year}{2014}).

\bibitem{johansson2012qutip}
\bibinfo{author}{Johansson, J.~R.}, \bibinfo{author}{Nation, P.~D.} \&
  \bibinfo{author}{Nori, F.}
\newblock \bibinfo{title}{Qutip: An open-source python framework for the
  dynamics of open quantum systems}.
\newblock \emph{\bibinfo{journal}{Comput. Phys. Commun.}}
  \textbf{\bibinfo{volume}{183}}, \bibinfo{pages}{1760--1772}
  (\bibinfo{year}{2012}).

\bibitem{virtanen2020scipy}
\bibinfo{author}{Virtanen, P.} \emph{et~al.}
\newblock \bibinfo{title}{Scipy 1.0: fundamental algorithms for scientific
  computing in python}.
\newblock \emph{\bibinfo{journal}{Nat. Methods}} \textbf{\bibinfo{volume}{17}},
  \bibinfo{pages}{261--272} (\bibinfo{year}{2020}).

\end{thebibliography}

\end{document}


\maketitle

\singlespacing

\section{Experimental configuration}

\subsection{Detailed experimental setup}
\begin{figure}[h!]
	\centering
    \includegraphics{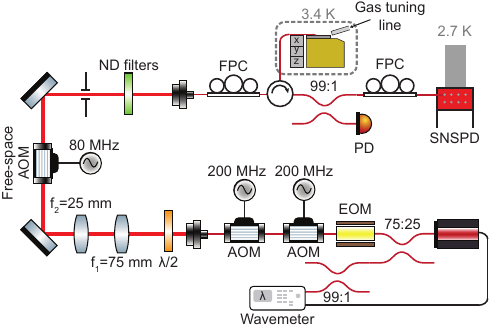}
    \caption{\textbf{Experimental setup.} A more detailed version of the experimental setup shown in Fig. 2a of the main text.
    }
	\label{fig:setup}
\end{figure}

In this section we give a more detailed description of our experimental setup, shown in Fig.~S\ref{fig:setup}. The fiber-coupled telecom CW tunable laser (Toptica CTL 1320) is frequency stabilized via a wavemeter (Bristol Instruments 871A) with a repeatability of 1.7 MHz. The analog error signal generated from the wavemeter is fed into the laser to perform PID-enabled laser locking. Excitation pulses are generated through a series of acousto-optic modulators (AOMs) and an electro-optic modulator (EOM). The AOM set consists of two fast-switching (10 ns speed) fiber AOMs (AeroDiode 1310-AOM-2) and one slow-switching free-space AOM (ISOMET M1205-P80L-0.6), providing a total extinction ratio of 142 dB. 
The intensity-modulated EOM is driven with a RF signal to generate laser sidebands. The generated laser pulses pass through a series of neutral density (ND) filters to control the optical power before entering the fiber network towards the devices.


The laser pulses pass through a polarization insensitive fiber circulator (Newport F-CIR-13-P-FP, insertion loss = 0.45 dB) before reaching the devices via the angle-polished fiber. The fiber is polished using an automatic fiber polishing machine (ULTRA TEC, NANOpol) at an angle of 37.97$^\circ$ for optimal coupling with the GC. The device is situated on the cold finger of a closed-cycle cryostat (Montana Cryostation s50) at a temperature of $T =$ 3.4 K. The cryostat has a magneto-optical component, which can apply a single-axis magnetic field up to 400 mT. The angle-polished fiber is mounted on a three-axis nanopositioner (Attocube ANPx101/LT, ANPz102/LT) to optimize its coupling to the GC. The T-center emission signal is routed by the circulator and sent into the superconducting nanowire single photon detector (SNSPD, Single Quantum EOS 210CS), which operates at a base temperature of $T =$ 2.7 K. A small fraction of the emission signal is sent to an InGaAs photodiode (PD, Newport 2053-FC) for monitoring the cavity spectrum. The SNSPD is biased using a waveform generator (Rigol DG832), and the voltage pulses from the SNSPD are counted using a time tagger (Swabian Instruments, Time Tagger 20). 

\subsection{Cavity tuning}

\begin{figure}[h!]
    \centering
    \includegraphics{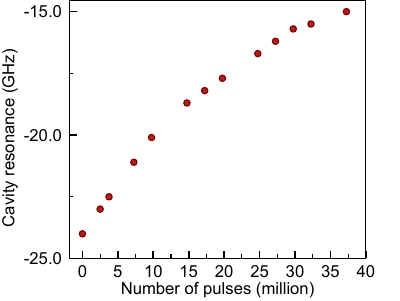}
    \caption{
    \textbf{Blue tuning of the cavity resonance.} The cavity resonance frequency is blue tuned by sending in optical pulses. The pulses are generated and controlled the same way as in the main experiments. 
    }
    \label{fig:blueTuning}
\end{figure}

The cavity resonance blue shifts about 9.8 nm when the device is cooled down from the room temperature to $T = 3.4$~K. Gas tuning \cite{Mosor2005} is implemented to tune the PC cavity resonance to match with the single T center's optical transition. We supply clean dry N$_2$ gas through a gas line inside the cryostat pointing to the device. Nitrogen ice then forms on the device surface, causing a red shift of the cavity resonance up to about 5 nm (850 GHz). To increase the tuning accuracy, we realize blue tuning of the cavity via optical pulses with high optical power to desorb the nitrogen ice. We send a 50\% duty cycle pulse train consisting of 1 $\mu$s wide optical pulses with 14.61 $\mu$W power at $\lambda_0 =$ 1326 nm to the cavity. Depending on the number of pulses, we can fine control the amount of blue tuning (Fig.~S\ref{fig:blueTuning}). The blue tuning process has an tuning accuracy $\sim$ 0.1 GHz. The condensation of the background gas causes a constant red shift of the cavity resonance at a rate of $\sim$ 10 MHz/min. We apply blue tuning in between experiments to compensate the background drift.

\subsection{Photon collection efficiency} \label{collectefficiency}

The system photon collection efficiency ($\eta_\text{sys}$), which represents the probability of a single photon emitted to the cavity being registered by the detector, is determined by 
$\eta_\text{sys}=\eta_\text{cav}\eta_\text{GC}\eta_\text{path}\eta_\text{SNSPD}$.
Here $\eta_\text{cav}= \kappa_\text{wg}/(\kappa_\text{wg}+\kappa_\text{sc})$ is the fraction of the photons in the cavity that couple to the waveguide, with $\kappa_\text{wg}$ and $\kappa_\text{sc}$ as the waveguide and scattering loss channels of the cavity, respectively; $\eta_\text{GC}$ is the one-way GC coupling efficiency at $\lambda_0=1326$ nm; $\eta_\text{path}$ is the transmission of the optical fiber network between the angle-polished fiber and the SNSPD; and $\eta_\text{SNSPD}$ is the SNSPD detection efficiency. A cavity efficiency $\eta_\text{cav}=0.358$ is estimated by measuring the ratio ($R$) of cavity reflection level on- and off-resonance using the relation $R=(1-2\eta_\text{cav})^2$ and assuming the cavity is under-coupled. We thus can determine the system photon collection efficiency to be $\eta_\text{sys}=0.091$ for the device that contains the single T center in the cavity ($\eta_\text{GC}=0.461$, $\eta_\text{path}=0.786$, $\eta_\text{SNSPD}=0.703$). For the cavity-coupled single T center analyzed in the main text, we observe an excited state 
Purcell factor $P_t=5.88$ when the cavity is on-resonance. The probability of the T center decaying into the resonant cavity mode is given by $\beta = P_t/(P_t+1) = 0.855$. 

Due to technical difficulties related to the SNSPD switching-on time ($\sim40$ ns) and free-space AOM slow switching ($\sim$ 150 ns), there exists a finite time window of $t_0 = 170$ ns in which the fluorescence from the single T center cannot be collected. Depending on the single T center lifetime $\tau_\text{exc}$, the amount of T center fluorescence that can be collected will be determined as $e^{-t_0/\tau_\text{exc}}$. Future improvement of the AOM switching time can decrease this time window down to the limit of the SNSPD switching time under the gating mode. When the cavity-coupled single T center emission is saturated, which corresponds to an incoherent excitation probability of 0.5, we are expecting a count level of 0.5$\eta_\text{sys}\beta e^{-t_0/\tau_\text{exc}}=0.011$. This prediction matches with the observed saturation count level of 0.01 photon per excitation pulse.

\section{Silicon photonic device design and fabrication}

\subsection{Cavity design and Purcell factor}

\begin{figure}[h!]
    \centering
    \includegraphics{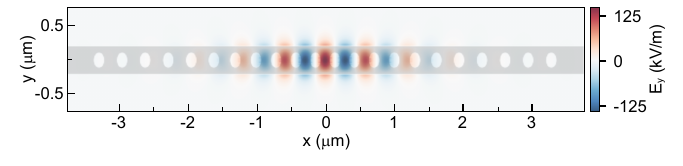}
    \caption{ \textbf{Cavity Mode profile.}
    Top view of the $E_y$ component of the electrical field in the cavity, at a $z$-cut in the middle of the silicon device layer ($z=110$ nm).
    } 
    \label{fig:EyDistribution}
\end{figure}

The PC cavity used in the experiments consists of an 1D array of elliptical holes, with the defect formed by parabolically decreasing the lattice constant in the center of the cavity \cite{chan2009optical}. The PC cavity design is carried out using the photonic calculation packages MPB \cite{johnson2001block} and MEEP \cite{oskooi2010meep}. The device layer thickness is $d_z = 220$ nm, and we use refractive indices for silicon and silicon dioxide of $n_\text{Si} = 3.505$ and $n_{SiO_2} = 1.466$, respectively. By maximizing the photonic bandgap while ensuring the correct symmetry, we obtain the final geometric parameters ($w, h_x, h_y, a_\text{mir}, a_\text{cav}) = (410, 140, 203, 334, 282)$ nm, where $w$ is the beam width, $h_x$ and $h_y$ are the two major axes of each PC hole, $a_\text{mir}$ and $a_\text{cav}$ are the lattice constant for reflector and cavity center regions, respectively. The simulated cavity field distribution is shown in Fig.~S\ref{fig:EyDistribution} with $N_\text{cav} = 14$ cavity holes at the center and $N_\text{mir} = 4$ mirror holes on each side. The mode volume of the cavity is calculated as $V_\text{mode} =$ 0.031 $\mu$m$^3$ $=0.172 (\lambda_0/n_\text{eff})^3$. To make an effective one-sided cavity for reflection measurements, we add 10 more mirror holes at the left end of the cavity. 

\begin{figure}[h!]
    \centering
    \includegraphics{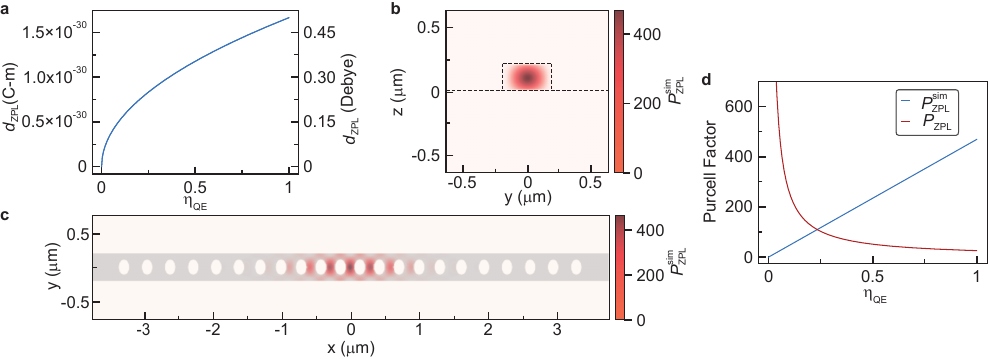}
    \caption{\textbf{T center ZPL dipole moment and Purcell factor.}  
    \textbf{a.} Dependence of T center ZPL transition dipole moment on the $\eta_\text{QE}$. 
    \textbf{b.} Simulated Purcell factor distribution in the $yz$-plane (at $x=0$) assuming $\eta_\text{QE} =1$. 
    \textbf{c.} Similar plot as panel \textbf{b} but now in the $xy$-plane (at $z=110$ nm).
    \textbf{d.} Simulated and indirectly measured ZPL Purcell factor assuming different $\eta_\text{QE}$.
    }
    \label{fig:PurcellDistribution}
\end{figure}

The Purcell factor for the T center's zero phonon line (ZPL) transition can be calculated as $P_\text{ZPL}^\text{sim} = 4g(\vec{r})^{2}/(\tilde{\kappa} \Gamma_{0})$, where $\Gamma_{0} = 2\pi \times 169.3$ kHz is the spontaneous emission rate \cite{Bergeron2020PRXQuantum}, $g(\vec{r})=\textbf{|\textit{d}}_\text{ZPL}\textbf{E}(\vec{r})|/\hbar$ is the local coupling constant, and $\tilde{\kappa}/2\pi = 7.11$ GHz is the linewidth determined in the detuning-dependent cavity enhancement (Fig.~4b of the main text). 
The ZPL transition dipole moment $\textbf{\textit{d}}_\text{ZPL}$ can be determined using the following equation, including a local field correction \cite{dung2006local},

\begin{equation}
    (\text{DW}\eta_\text{QE})\Gamma_{0} = \gamma_\text{ZPL} = \left( \frac{3 n_\text{Si}^{2}}{2n_\text{Si}^{2} + 1} \right)^{2}
    \frac{n_\text{Si} \textbf{\textit{d}}_\text{ZPL}^2 \omega^3 }{3\pi \epsilon_{0} \hbar c^{3}},
    \label{eq:dipoleMoment}
\end{equation}

\noindent where DW $= 0.23$ is the Debye-Waller factor for the T center \cite{Bergeron2020PRXQuantum} and $\omega = 2\pi \times 226141.974$ GHz is the cavity-coupled single T center optical transition frequency.
Depending on the quantum efficiency $\eta_\text{QE}$, we will have different $\textbf{\textit{d}}_\text{ZPL}$ (Fig.~S\ref{fig:PurcellDistribution}a). In an ideal case of $\eta_\text{QE} = 1$, the ZPL dipole moment will be $\textbf{\textit{d}}_\text{ZPL} = 1.67 \times 10^{-30}$ C-m (0.50 Debye). The predicted ZPL Purcell factor distribution inside the silicon cavity are shown in Fig.~S\ref{fig:PurcellDistribution}b and Fig.~S\ref{fig:PurcellDistribution}c. At the cavity center, assuming perfect dipole alignment with the cavity polarization, we will have the maximum Purcell factor $P_\text{ZPL}^\text{sim} = 470$ and the coupling $g = 2\pi \times 376$ MHz. From the measured cavity-enhanced decay, we can also predict the $P_\text{ZPL} = P_t/(\text{DW}\eta_\text{QE})$. Due to the suboptimal positioning of the single T center inside the cavity and imperfect dipole alignment with the local cavity polarization, we will have $P_\text{ZPL} \leq P_\text{ZPL}^\text{sim}$. This enables us to extract a lower bound of the quantum efficiency of $\eta_\text{QE} \geq 23.4 $\% (Fig. S\ref{fig:PurcellDistribution}d).


\subsection{Grating coupler design}

The light is coupled into and out-of the devices via a 2D subwavelength GC, shown in Fig.~1b of the main text. The GC consists of 20 periods of fully-etched PC holes with a triangular lattice \cite{liu2010high,ding2013ultrahigh}, and a width set to 13.1 $\mu$m to maximize the mode overlap along the $y$-direction between the GC diffracted light and a SMF-28 fiber (mode field diameter = 9.2 $\mu$m). A uniform grating structure does not provide a good mode overlap with the fiber Gaussian mode \cite{halir2015waveguide}. Instead we turn to an apodized structure to fine tune the diffraction strength at each period to maximize the mode overlap.

\begin{figure}[h!]
    \centering
    \includegraphics{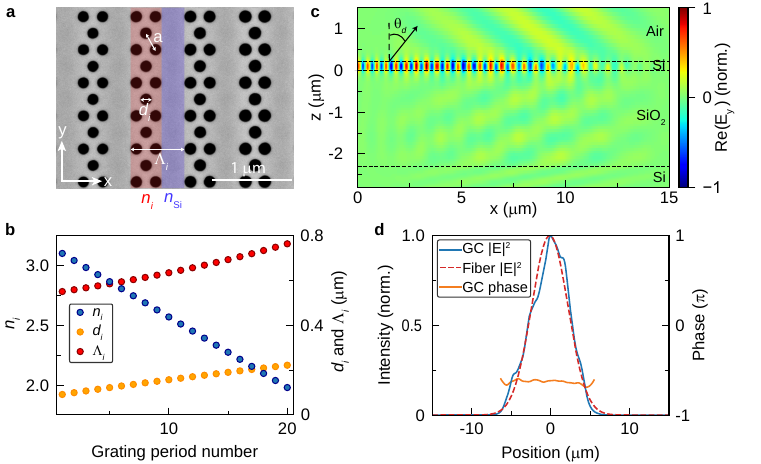}
    \caption{\textbf{Subwavelength grating coupler design.}
    \textbf{a.} Scanning electron microscope image of a section of the GC. Structural parameters of the GC are labeled on the image. Here, $n_i$, $\Lambda_i$, $d_i$ and $a$ are the effective index, period, PC hole diameter and PC hole pitch in each grating period, respectively. The duty cycle of the grating structure is kept constant at $1-(a + d_i)/\Lambda_i = 0.4$.
    \textbf{b.} Parameters of $n_i$, $\Lambda_i$ and $d_i$ for the optimized apodized grating structure. 
    \textbf{c.} Simulated electric field ($E_y$) radiated by the GC in the $xz$-plane (at $y=0$). $\theta_d$ denotes the diffraction angle.
    \textbf{d.} Simulated intensity and phase profiles of the GC diffracted field along ($-\cos(\theta_d)\hat{x} + \sin(\theta_d)\hat{z}$) direction,
    and the mode profile (red dashed line) of a SMF-28 fiber. 
    }
    \label{fig:gratingcoupler}
\end{figure}

Three-dimensional Lumerical (Ansys) finite-difference time-domain (FDTD) simulations are used to optimize the coupling efficiency of the GC, $\eta_\text{GC} = D\times O_\text{GC}^x O_\text{GC}^y$ at $\lambda_0 = 1326$nm, where $D$ is the directionality of the GC, quantifying the ratio of the light diffracted towards the fiber; $O_\text{GC}^x$ and $O_\text{GC}^y$ are the mode overlap between GC diffracted light and the fiber mode along ($-\cos(\theta_d)\hat{x} + \sin(\theta_d)\hat{z}$) and $\hat{y}$ directions, respectively. To improve the mode overlap $O_\text{GC}^x$, the effective index $n_i$ (Fig.~S\ref{fig:gratingcoupler}a) for each grating period is linearly apodized by changing the diameter ($d_i$) of the PC holes \cite{ding2013ultrahigh} while keeping the lattice pitch constant ($a = 239$ nm). The period $\Lambda_i$ is adjusted accordingly to ensure a constant diffraction angle $\theta_d$ at different grating periods. To fully utilize the SOI structure, the thickness of the buried oxide layer is customized as $d_\text{BOX} = 2.3$ $\mu$m to create constructive interference in the GC diffraction and thus increase the directionality $D$. The detailed procedure for optimizing the GC efficiency can be found in Ref. \cite{halir2009waveguide, chen2021hybrid}.

The optimized geometric parameters are summarized in Fig.~S\ref{fig:gratingcoupler}b, where the initial grating period and effective index are $\Lambda_1 = 550$ nm and $n_1 = 3.1$, respectively. The index change between neighbouring period is $\Delta n = 0.059$. The electric field distribution of the GC diffracted field is shown in Fig.~S\ref{fig:gratingcoupler}c with a diffraction angle $\theta_{d}=20.88^{\circ}$ and a directionality $D=76.3 \%$. The fiber polishing angle can be calculated as $\alpha = [90^\circ - \sin^{-1}(\sin \theta_d/n_\text{fiber})]/2 = 37.97^{\circ}$ ($n_\text{fiber} = 1.4676$). The mode overlap between the GC diffracted field and the fiber Gaussian mode are $O_\text{GC}^x = 98.6 \% $ (Fig.~S\ref{fig:gratingcoupler}d) and $O_\text{GC}^y = 98.9 \% $. After taking into account the transmission of the linearly tapered waveguide (95.9\%) and the glass-air interface reflection loss (3.6\%), we calculate a maximum one-way coupling efficiency $\eta_\text{GC}=68.8\%$.

During fabrication, we add a constant change on all the PC hole sizes in the GC to tune the GC spectrum center to about 1331 nm at room temperature. After cooling down, we observe $\sim$ 5 nm blue shift for the GC spectrum. The typical GC efficiency at room temperature is $\eta_\text{GC} \sim 60\%$ at the GC spectrum center
in a measurement setup with full six degrees of freedom ($x$, $y$, $z$, pitch, yaw, rotation) control of the fiber. After fiber gluing (Stycast 2850FT, Catalyst 9) and mounting onto the three-axis ($x$, $y$, $z$) piezo nanopositioner, $\eta_\text{GC}$ typically drops by $\sim$ 5\%, which is likely due to the change of the pitch angle of the fiber with respect to the sample.


\subsection{Nanofabrication process} \label{fabProcess}

\begin{figure}[h!]
    \centering
    \includegraphics{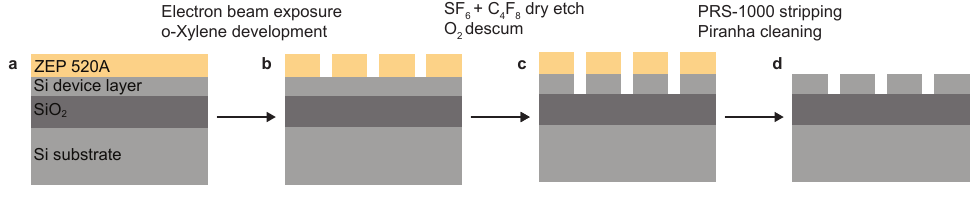}
    \caption{\textbf{Nanofabrication process flow.}
    \textbf{a.} Cross-sectional view of a SOI sample with an electron beam resist layer on top. 
    \textbf{b.} The nanophotonic circuit pattern forms on the resist layer after electron beam writing and development. 
    \textbf{c.} Reactive ion etching transfers the pattern to the silicon device layer, followed by oxygen descum. 
    \textbf{d.} Resist stripping and piranha cleaning. 
    }
    \label{fig:nanoFab}
\end{figure}

Our nanofabrication process is outlined in Fig.~S\ref{fig:nanoFab}. All of the nanophotonic devices are fabricated on SOI samples (WaferPro). The SOI has a $220\pm10$ nm float zone grown P-type device layer with resistivity $\geq$ 1000 $\ohm\cdot$cm. The buried oxide has a customized thickness of 2.3 $\mu$m for maximizing the GC coupling efficiency, and the handling layer has a thickness of 725 $\mu$m.

We spin coat 400 nm electron beam (ebeam) resist ZEP520A (Zeon Specialty Materials Inc.) onto 9$\times$9 mm$^2$ SOI chips and bake at 170 $^\circ$C for 5 mins. The sample is exposed using an Elionix ELS-G100 ebeam writer with a dosage of 225 $\mu$C/cm$^2$, and subsequently developed in o-xylene at room temperature for 90 seconds and rinsed in isopropanol for 20 seconds. The pattern is then defined on the resist layer, which acts as the etching mask. The sample etching is performed using an inductively coupled plasma (ICP) reactive ion etcher (Oxford Plasmalab System 100/ICP 180) with SF$_{6}$/C$_{4}$F$_{8}$ gases. The sample is kept at 0$^\circ$~C during the etching process. After etching, the sample goes through a series of processes including oxygen plasma descum, dicing (into 4.5$\times$4.5 mm$^2$), resist stripping, and piranha cleaning before being transferred into the cryostat for measurements.

\subsection{Cavity quality factor vs. T center generation}

To determine the potential influence of the T center generation process on the cavity quality factor, we fabricate devices on both implanted and unimplanted SOI samples. The implanted sample has an equal fluence of $5\times10^{13}$ cm$^{-2}$ for $^{12}$C and $^1$H implantation and has gone through corresponding thermal annealing steps, following the procedure discussed in Section \ref{ionimplantprotocol}. The two samples' respective ebeam writings are done under the same exposure condition using the same pattern. The ebeam writing tool's conditions (e.g., beam focus, stigmation, beam current) have been kept the same for the two writings to the best of our control. The two samples are then placed on the same carrier wafer and etched simultaneously during the etching step.

\begin{figure}[h!]
    \centering
    \includegraphics{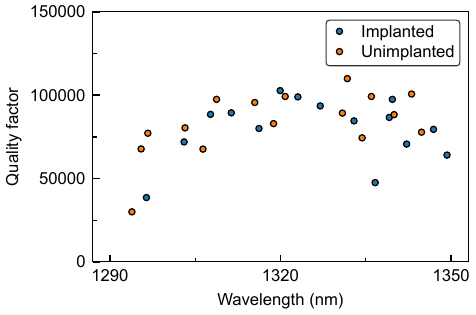}
    \caption{\textbf{Cavity Quality factors.} 
    Measured cavity quality factors ($Q$) for devices fabricated on the implanted and the unimplanted samples. No significant difference in the device $Q$ performance is observed between the two samples.
    }
    \label{fig:QualityVsImplant}
\end{figure}

To ensure finding cavities that can be gas tuned into resonance with the T center optical transitions, we sweep the hole sizes in the PC cavities, which changes the cavity resonance frequency. The measured quality factors for the two samples are shown in Fig.~S\ref{fig:QualityVsImplant}, revealing $Q= (8.0\pm1.8)\times10^4$ and $Q= (8.4\pm1.8)\times10^4$ for implanted and unimplanted SOI samples, respectively. This confirms that the T center generation process has a minimal effect on the cavity quality factor up to an implantation fluence of $5\times10^{13}$ cm$^{-2}$.

\section{T center generation via ion implantation}

\subsection{Generation protocol} \label{ionimplantprotocol}

T centers shown in this work are generated via uniform ion implantation method following a published procedure \cite{MacQuarrie2021}. We use an equal fluence for $^{12}$C and $^{1}$H during the ion implantation processes (II-VI Coherent Corp.).

Secondary ion mass spectroscopy (SIMS, Eurofins EAG) analysis of our SOI device layer before ion implantation shows that the native carbon and hydrogen densities are $\left[ \text{C} \right] = 3\times10^{17}$ cm$^{-3}$ and $\left[ \text{H} \right] \approx 5\times10^{17}$ cm$^{-3}$, respectively. We note that the hydrogen density is close to the SIMS tool limit for hydrogen species ($1\times10^{17}$ cm$^{-3}$).

\begin{figure}[h!]
    \centering\includegraphics{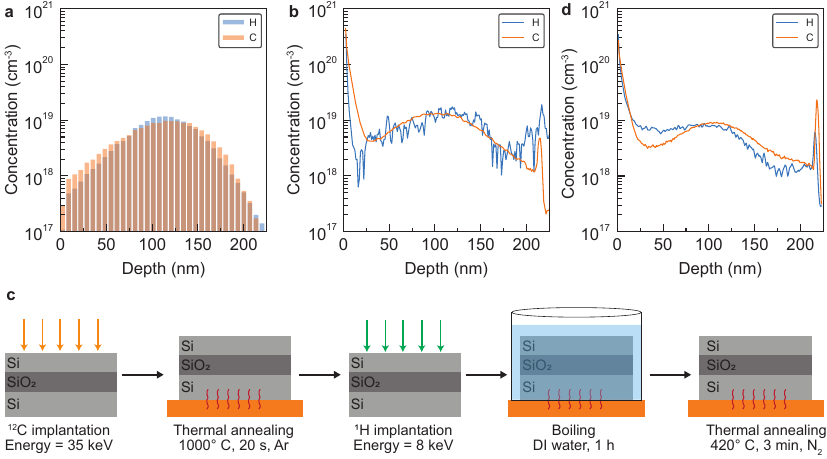}
    \caption{\textbf{T center generation process} \textbf{a.} SRIM simulation results for 35 keV $^{12}$C and 8 keV $^{1}$H implantation with an equal fluence of $1\times10^{14}$ cm$^{-2}$.
    \textbf{b.} SIMS analysis of a test sample that goes through the same implantation conditions simulated in panel \textbf{a}. No thermal annealing process is performed for this test sample. 
    \textbf{c.} Process flow for T center generation. 
    \textbf{d.} SIMS analysis of a SOI sample that goes through the T center generation process. The implantation fluence is $1\times10^{14}$ cm$^{-2}$ for both $^{12}$C and $^{1}$H.
    }
    \label{fig:Tcenter_generation}
\end{figure}

We use the stopping and range of ions in matter (SRIM) software package \cite{Ziegler2010} to estimate the ion implantation energy necessary for the carbon and hydrogen ions to end up in the middle of the SOI device layer ($d_\text{depth} = 110$ nm). We determine the implantation energy for $^{12}$C and $^{1}$H to be 35 keV and 8 keV, respectively, for an ion implantation direction of 7$^{\circ}$ from the normal incidence to avoid channeling effect in silicon \cite{Cho1985}. With an equal fluence of $1\times10^{14}$ cm$^{-2}$ for both $^{12}$C and $^{1}$H, the simulation suggests a peak density of $1\times10^{19}$ cm$^{-3}$ for both species in the middle of the device layer (Fig.~S\ref{fig:Tcenter_generation}a). The FWHM linewidth of the distribution is 89.3 $\pm$ 0.8 nm and 79.8 $\pm$ 0.6 nm for $^{12}$C and $^{1}$H implantation, respectively. To verify this, we implement sequential ion implantation test runs for $^{12}$C and $^{1}$H (with equal fluence of $1\times10^{14}$ cm$^{-2}$) without any thermal annealing process. The SIMS measurement of this test sample (Fig.~S\ref{fig:Tcenter_generation}b) matches well with the SRIM simulation results.

The detailed procedure to generate T centers is shown in Fig.~S\ref{fig:Tcenter_generation}c. We perform $^{12}$C ion implantation at 7$^{\circ}$ direction, 35 keV energy, followed by rapid thermal annealing (RTA) at 1000 $^{\circ}$C for 20 seconds under an Argon background to repair lattice damage and substitute the implanted carbon \cite{higginbottom2022optical}. Next, a second round implantation of $^{1}$H at 7$^{\circ}$ direction, 8 keV energy is performed. After the two implantation steps, we boil the sample for 1 hour in DI water, followed by a second RTA process at 420 $^{\circ}$C for 3 minutes with a N$_2$ background. We implant a SOI sample following the above procedures with an equal fluence of $1\times10^{14}$ cm$^{-2}$ for $^{12}$C and $^{1}$H. The SIMS analysis (Fig.~S\ref{fig:Tcenter_generation}d) reveals carbon and hydrogen densities of $8.87\times10^{18}$ cm$^{-3}$ and $7.95\times10^{18}$ cm$^{-3}$, respectively, in the middle of the device layer.

For sample A and B measured in the main text, the implantation fluences are $5\times10^{13}$ cm$^{-2}$ and $7\times10^{12}$ cm$^{-2}$, respectively. 
We use the PLE spectra measured when the cavity is far-detuned from multiple devices to estimate the T center density. Due to spectral congestion, the T center peaks near the center of the inhomogeneous distribution are not well resolved. We turn to use the PLE spectrum area to estimate T center density, where we calculate the ratio between the area under the full PLE spectrum and the average area under a well isolated T center peak at the tail of the inhomogeneous distribution. 
We estimate an average number of $\sim$100 and $\sim$30 T centers per device for sample A and B, respectively. These amounts of T centers correspond to a T center density of $\sim1 \times10^{12}$ cm$^{-3}$ and $\sim 3 \times10^{11}$ cm$^{-3}$ for sample A and B, respectively, assuming the majority of the observed T centers are in the taper waveguide region. Further optimization of the ion implantation fluence ratio ($^{12}$C:$^1$H) and the thermal treatment are underway to improve the T center generation yield.

\subsection{Nuclear spin density analysis} \label{nuclearspin}

The nuclear spin bath around the single T center can create magnetic noises and cause dephasing. Here, we analyze potential atomic species with non-zero nuclear spins. To gauge the average distance between neighbouring nuclear spins of the same atomic species, we utilize the following formalism to calculate. We define a probability distribution $w(r)$ to describe the probability of finding a neighbouring nuclear spin within distance $r$, which satisfies,

\begin{equation}
    w(r) dr = \left(1 - \int_{0}^{r} w(r') dr' \right)(4\pi r^{2} dr) \rho,
    \label{probDistribution}
\end{equation}

\noindent where $\rho$ is the nuclear spin volume density. Solving Eq. \ref{probDistribution} will lead to a probability distribution function,

\begin{equation}
    w(r) = 4\pi\rho r^2 \text{exp}\left(-\frac{4\pi\rho}{3} r^3\right).
\end{equation}

\noindent The average distance between the same-species nuclear spins ($d_{nn}$) is then derived as,

\begin{equation}
    d_{nn} = \int_{0}^{\infty} w(r)r dr = \Gamma(4/3)\left(\frac{4\pi\rho}{3}\right)^{-\frac{1}{3}} \approx 0.554 \rho^{-\frac{1}{3}},
\end{equation}

\noindent where $\Gamma(z)=\int_{0}^{\infty} t^{z-1}e^{-t}dt$ is the Gamma function.

The table below shows major nuclear spin species inside the sample B (C:H = 1:1, $7\times10^{12}$ cm$^{-2}$ implantation fluence).
As a reference to the density shown below, the silicon lattice has an atomic density of $5.0 \times10^{22}$ cm$^{-3}$. The boron density is estimated using the SOI device layer resistivity \cite{thurber1980resistivity}. 

\begin{center}
    \begin{tabular}{ |c||c|c|c|c|c|  }
     \hline
        \rule{0pt}{2.5ex} & $^{29}$Si ($I$=1/2) & $^1$H ($I$=1/2)& $^{13}$C ($I$=1/2) & $^{11}$B ($I$=$-$3/2) & $^{10}$B  ($I$=3)\\
     \hline
        \rule{0pt}{2.5ex} Density (cm$^{-3}$) & $2.34\times10^{21}$  & $5.56\times10^{17}$ & $6.58\times10^{15}$ & $1.07\times10^{13}$ & $2.61\times10^{12}$ \\
     \hline
        \rule{0pt}{2.5ex} Average separation (nm)  & 0.42   & 6.7 &  29.6  & 251.4 & 402.4   \\
     \hline
    \end{tabular}
\end{center}

The major nuclear spin species is $^{29}$Si, which can be isotopically purified to reach $\sim$ ppm level, therefore lowering the density of $^{29}$Si to $\sim 10^{16}$ cm$^{-3}$. 
As hydrogen is part of the T center, its density can't be decreased without compromising the T center density, which necessitates trade off between them. With a proper density, environmental hydrogen can be utilized as extra nuclear spin quantum memories \cite{uysal2023coherent}. 




\section{Waveguide-coupled T centers}

\begin{figure}[h!]
	\centering\includegraphics{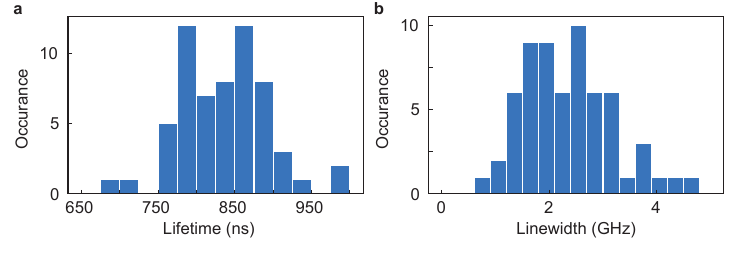}
    \caption{
    \textbf{Statistics of waveguide-coupled T centers.}
    \textbf{a.} Excited-state lifetime.
    \textbf{b.} FWHM linewidth. 
   }
	\label{fig:histograms}
\end{figure}

With the cavity far-detuned from the scan range, we probe the T centers that are most likely in the taper waveguide region. We note that the spectrum (Fig. 2c of the main text) measured in this case can also include contributions from T centers located in the GC region, but with less quantity as the volume of the GC is $\sim$10 times smaller than that of the taper waveguide. Figure S\ref{fig:histograms}a and S\ref{fig:histograms}b summarize the fluorescence lifetime and linewidth from resolvable T center peaks across different devices from sample A and B. The statistical analysis reveals an average lifetime of $\tau_\text{exc}=836.8 \pm 57.3 $ ns and FWHM linewidth of $\Gamma=2.40 \pm0.86 $ GHz. 

\begin{figure}[h!]
	\centering\includegraphics{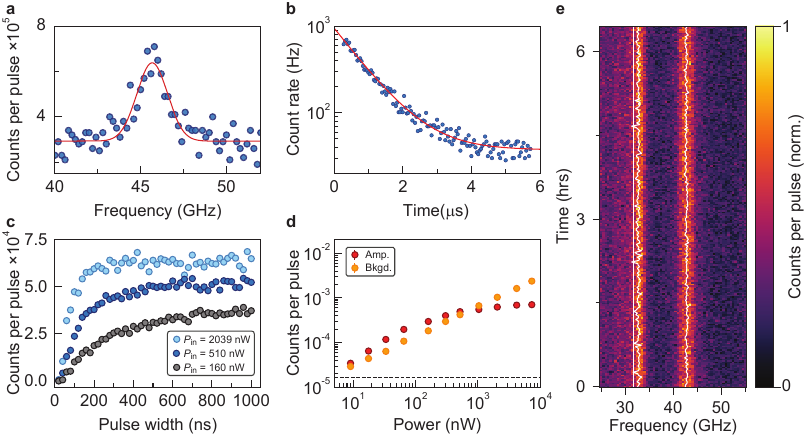}
    \caption{
    \textbf{Waveguide-coupled T center characterizations.} 
    \textbf{a.} PLE spectrum of a waveguide-coupled T center at $P_\text{in} = 9.12$ nW. Red line is the Gaussian fitting.  
    \textbf{b.} Time-resolved fluorescence decay for the same T center at $P_\text{in} = 302.68$ nW. 
    \textbf{c.} Optical Rabi measurement at different powers. 
    \textbf{d.} Saturation curve with an excitation pulse width of 600 ns. The dashed black line is the contribution from the detector dark counts. Here the collection window misses the first 300 ns of the fluorescence decay.
    \textbf{e.} Spectral diffusion of two waveguide-coupled T centers at an excitation power $P_\text{in}=60.93$ nW. White solid line shows the Gaussian fitted peak center at each iteration (with a duration of 102.5 seconds). The T center occasionally shows spectral jumps, causing the peak position difference between two measurements of the same T center located at 45.67 GHz in panel \textbf{a} and 42.78 GHz in panel \textbf{e}.
    }
\label{fig:waveguide_Tcenter}
\end{figure}

The average $\tau_\text{exc}$ is slightly shorter than the bulk lifetime of $1/\Gamma_0 = 940$ ns \cite{Bergeron2020PRXQuantum}, which may come from a finite Purcell enhancement due to the waveguide and/or more unwanted nonradiative decays in the nanophotonic structure compared with the bulk silicon. The linewidth distribution is similar to previously reported measurements \cite{higginbottom2022optical}. We note that a large power ($P_\text{in} \sim \mu$W) is used when measuring the PLE spectrum for waveguide-coupled T centers, therefore power broadening is expected in the measured $\Gamma$. The observed cavity-coupled T center, unfortunately, has a linewidth at the higher end of this distribution. Further investigation is needed to verify whether single T centers inside the cavity have a systematically larger linewidth compared with those in the SOI waveguides or bulk silicon.


We investigate the optical properties of a specific waveguide-coupled T center located at 45.67 GHz in sample B, which has a low-power linewidth $\Gamma = 2.10 \pm 0.18$ GHz (Fig.~S\ref{fig:waveguide_Tcenter}a) and a fluorescence lifetime of $\tau_\text{exc}=838.2 \pm 8.1$ ns (Fig.~S\ref{fig:waveguide_Tcenter}b). 
Due to the fast dephasing, this waveguide-coupled T center is incoherently driven to saturation without any Rabi oscillations (Fig.~S\ref{fig:waveguide_Tcenter}c). The emission of this waveguide-coupled T center shows typical saturation behavior (Fig.~S\ref{fig:waveguide_Tcenter}d) with a saturated count level of $C_p^\text{sat} = 7\times 10^{-4}$ photon counts per pulse. Using a similar procedure discussed in Section \ref{collectefficiency}, we can define a system collection efficiency for a waveguide-coupled T center as $\eta_\text{sys,wg} = \eta_\text{wg}\eta_\text{col}\eta_\text{path}\eta_\text{SNSPD}=0.0785\times\eta_\text{wg}$, where $\eta_\text{wg}$ is the efficiency for the T center emission being coupled to the waveguide mode. The $\eta_\text{col}=0.142$ is the fraction of the T center ZPL and PSB outcoupled by the GC, estimated using the T center's DW factor and PSB spectrum under resonant PL excitation reported in literature \cite{Bergeron2020PRXQuantum}.
The saturation count level can thus be calculated as $C_p^\text{sat}=0.5e^{-t_{0}/\tau_\text{exc}}\eta_\text{QE}\eta_\text{sys,wg}$, which let us extract the waveguide efficiency $2.55\% \leq \eta_\text{wg} \leq 10.90\%$. 
Similarly as in Fig.~3e of the main text, we monitor the stability of two waveguide-coupled T centers
for up to 6 hours (Fig.~S\ref{fig:waveguide_Tcenter}e), revealing a spectrum-center distribution of $32.73 \pm 0.31$ GHz and $42.78 \pm 0.18$ GHz for the two T centers, respectively.




\section{Cavity-coupled single T center}


\subsection{Optical Rabi}\label{subsec:cavityTRabi}

\begin{figure}[h!]
	\centering \includegraphics{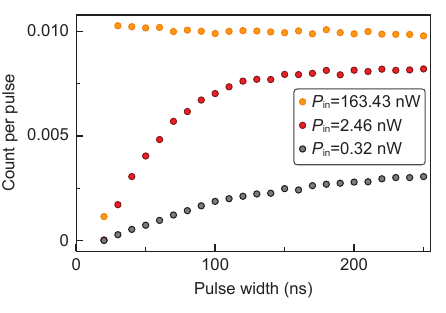}
    \caption{\textbf{Optical Rabi measurement for the cavity-coupled single T center.} The curve is measured when the cavity is on resonance.
    The counts per pulse is background corrected.
    }
	\label{fig:cavity_Tcenter}
\end{figure}

Due to the large dephasing rate, we have not been able to drive coherent optical Rabi oscillations of the cavity-coupled single T center (Fig.~S\ref{fig:cavity_Tcenter}) up to an excitation power of $P_\text{in}=163.43$ nW, which corresponds to a Rabi frequency $\Omega = 2\pi \times 586$ MHz. 
We use a pulse width of 900 ns to ensure incoherent driving for the cavity-coupled single T center in different measurements.


\vspace{-10pt}
\subsection{Second-order autocorrelation $g^{(2)}(0)$ analysis}

Higher $g^{(2)}(0)$ is observed at increasing excitation powers during the saturation measurements (Fig.~3d of the main text). The emission from waveguide-coupled T centers (e.g., those from the taper waveguide and the GC) contributes to the accidental $g^{(2)}(0)$ coincidences, which leads to a dependence of $g^{(2)}(0)$ on the signal-to-noise ratio (SNR, $A$). We note that the detector dark counts (3 Hz) have much less influence on the $A$. The $g^{(2)}(0)$ can be described as \cite{dibos2018atomic},

\begin{equation}
    g^{(2)}(0)= \frac{2A+1}{(A+1)^2}.
    \label{eq:g2_lim}
\end{equation}



\begin{figure}[h!]
	\centering \includegraphics{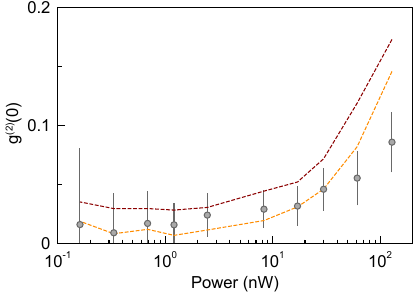}
    \caption{\textbf{Second-order autocorrelation.} $g^{(2)}(0)$ as a function of $P_\text{in}$ for the cavity-coupled single T-center at $\Delta_{ac}=0$. Dashed lines show the SNR-limited $g^{(2)}(0)$ calculated by Eq. \ref{eq:g2_lim} with the SNR estimated using a background from the Gaussian fitting (red) and from the average spectrum intensity at a region away from the T center peak (orange).
    }
	\label{fig:g2(0)}
\end{figure}

\noindent The SNR, $A$, is first extracted using the Gaussian-fitted amplitude and background for each PLE spectrum. Predicted $g^{(2)}(0)$ based on Eq.~\ref{eq:g2_lim} (red dashed line in Fig.~S\ref{fig:g2(0)}) surpasses the experimental results, which is due to the inaccurate estimation of the PLE spectrum background. To better estimate the $A$, the background is obtained by using the average spectrum intensity at a region away from the T center peak. The updated $g^{(2)}(0)$ prediction (orange dashed line in Fig.~S\ref{fig:g2(0)}) shows a good match with the data, confirming the $g^{(2)}(0)$ is limited by the SNR.


\subsection{Bunching behavior of the autocorrelation}

\begin{figure}[h!]
    \centering
    \includegraphics{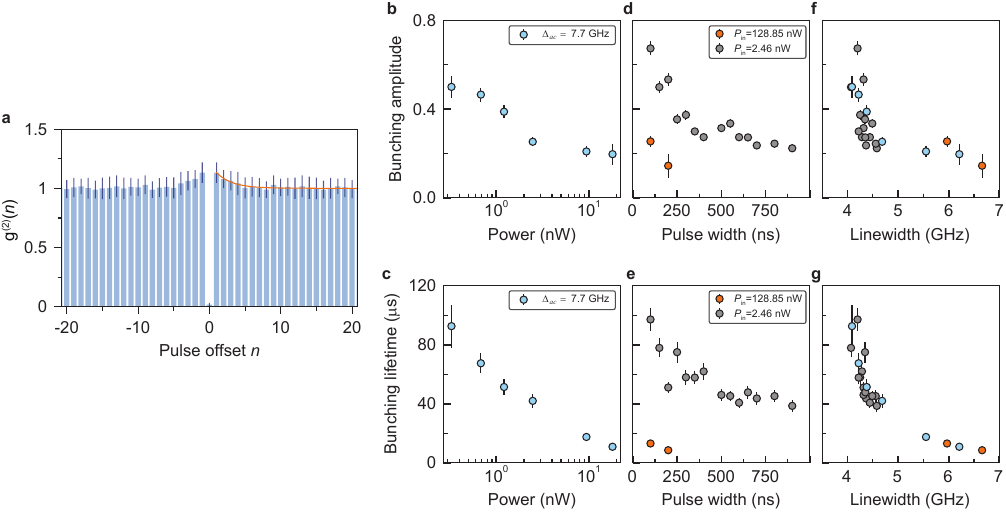}
    \caption{
    \textbf{Autocorrelation bunching behavior.}
    \textbf{a.} A representative $g^{2}(n)$ plot that shows bunching feature with $\Delta_{ac}=7.7$ GHz, under an excitation power of $P_\text{in} = 9.37$ nW with an excitation pulse width of 900 ns. The single-exponential decay (orange line) reveals a bunching lifetime of $17.5 \pm 2.3$ $\mu$s. Here the bunching lifetime is presented as $nT_\text{seq}$, where $ T_\text{seq} = 8$ $\mu$s is the length of each pulse sequence. 
    Panels \textbf{b} and \textbf{c} show the bunching amplitude and lifetime under different input powers at $\Delta_{ac}=7.7$ GHz with a pulse width of 900 ns. 
    Panels \textbf{d} and \textbf{e} show the bunching amplitude and lifetime under two excitation powers at different pulse widths ($\Delta_{ac}=-7.7$ GHz).
    Panels \textbf{f} and \textbf{g} show the summarized bunching amplitude and lifetime results from panel \textbf{b} $-$ \textbf{e}, but now using the corresponding PLE spectrum linewidth as the $x$-axis.
    }
    \label{fig:g2Bunching}
\end{figure}

In this section, we discuss the bunching features in the autocorrelation measurements of the cavity-coupled single T center. A typical $g^{2}(n)$ with bunching behavior is shown in Fig.~S\ref{fig:g2Bunching}a. We measure the autocorrelation under different experimental conditions ($P_\text{in}$, pulse width, $\Delta_{ac}$) to investigate factors that may affect the bunching feature (Fig.~S\ref{fig:g2Bunching}b$-$e). Qualitatively, the bunching is less pronounced when the cavity-coupled single T center is driven stronger (e.g., under higher power, wider pulse width, and smaller detuning).


The bunching in the autocorrelation measurements for a single T center at zero magnetic field can be the result of transient occupation of a dark state (e.g., a shelving state or a charge state)
or spectral diffusion \cite{sallen2010subnanosecond}. Similar bunching dynamics have been reported in single G centers in silicon due to a shelving sate \cite{redjem2020single}. However, such a shelving mechanism has not been demonstrated for single T centers to the best of our knowledge. The bound exciton excited states (e.g., TX$_1$, TX$_2$...) are likely having very short lifetimes due to the fast phonon relaxation. As the bunching decay lifetimes are at tens of $\mu$s time scales ($\gg$ fluorescence decay), we conclude that they can't originate from the bound exciton excited states. 
Charge state seems unlikely for the cavity-coupled single T center as no fluorescence intermittency is observed (Fig.~3e of the main text), and the count level keeps at saturation under a wide range of high powers (Fig.~3c of the main text). 

The above analysis leads us to believe the spectral diffusion is the main origin of the observed bunching feature. The cavity-coupled single T center indeed has a large spectral diffusion as shown in Section \ref{numericalmodeling}.
When summarizing all bunching lifetimes and amplitudes against PLE spectrum linewidth (Fig.~S\ref{fig:g2Bunching}f$-$g), we observe strong dependence unexpectedly. Further investigations are needed to illustrate the specific mechanisms involved here. Several noise sources can generate spectral diffusion, including the magnetic field noises due to the nuclear spin bath of $^{29}$Si and potential electrical field noises from silicon surface charge states or other impurities existing in the silicon. 

\subsection{Magnetic field test}

The electronic structure of a single T center consists of a ground state formed by an isotropic electron spin ($g$-factor $g_e=2.005$), and an excited state formed by an anisotropic hole spin \cite{Bergeron2020PRXQuantum}. Depending on the T center orientation with respect to the applied magnetic field $B$, the anisotropic hole spin has a $g$-factor ($g_h$) ranging from 1.609 to 3.457 when the $B$ field is along the silicon $\langle110\rangle$ direction. For $B$ field along the $\langle100\rangle$ direction, the $g_h$ values for the twelve possible T center orientational subsets reduce to $g_h = \{0.91, 2.55\}$ with multiplicities of 4 and 8, respectively \cite{MacQuarrie2021}.

\begin{figure}[h!]
    \centering
    \includegraphics{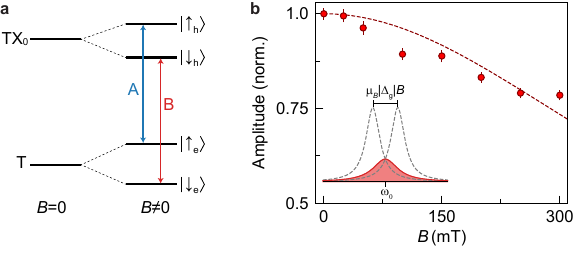}
    \caption{\textbf{Magnetic field test.} 
    \textbf{a.} Schematic of T center's electronic structure. With $B\neq0$, the doublet ground and TX$_0$ states will be split.
    \textbf{b.} PLE peak intensity decrease of the cavity-coupled T center ($\Delta_{ac} = 0$) at increasing magnetic field along silicon $[100]$ direction. The experimental data is fitted by Eq. \ref{eq:g-factor} (dashed line).
    }
    \label{fig:magneticfield}
\end{figure}

We apply a magnetic field along the silicon $[100]$ direction up to 300 mT aiming to split the cavity-coupled single T center's ZPL transition. To avoid optical pumping that may polarize the electron spin \cite{Bergeron2020PRXQuantum}, we use the EOM to generate RF sidebands ($f_0 \pm f_\text{RF}$) of the laser carrier ($f_0$) and drive the two spin conserving transitions simultaneously (transitions A and B in Fig. S\ref{fig:magneticfield}a). By scanning the $f_\text{RF}$, we expect to locate a fluorescence peak when $f_\text{RF} = \frac{1}{2} |\Delta_g| \mu_B B$, where $\mu_B$ is the Bohr magneton and $|\Delta_g| = |g_e - g_h|$ is the difference of the excited- and ground-state $g$-factors. We note that we have not been able to observe such a Zeeman splitting, which may be due to the insufficient splitting compared with the broad linewidth of the cavity-coupled single T center.

We then turn to the single-tone laser carrier scan. The optical pumping will cause electron spin polarization \cite{Bergeron2020PRXQuantum}, which reduces the PLE signal. However, the broad single T center linewidth enables the single frequency excitation to address both spin conserving transitions in their spectral overlap (inset in Fig.~S\ref{fig:magneticfield}b). As the separation between A and B transitions increases at higher $B$ field, the probability of driving both transitions decreases, which leads to spin polarization and the lowering of the PLE signal. We note that the T center position and its fluorescence lifetime remain the same across the whole magnetic field test, suggesting minimal influence from the coupling with the cavity. 

Assuming that the optical linewidth $\Gamma$ for transitions A and B are both equal to the linewidth for the cavity-coupled single T center at zero field ($\Gamma = 3.83$ GHz under $P_\text{in}=0.08$ nW), we can model the amplitude of the field-dependent PLE spectrum using the following equation \cite{MacQuarrie2021},

\begin{equation}
    I(B) = \frac{\Gamma^2}{\Gamma^2+(|\Delta_g| \mu_B B)^2}.
    \label{eq:g-factor}
\end{equation}

\noindent For simplicity, we have neglected the minor contributions from the spin-nonconserving transitions. By fitting the field dependent PLE spectrum amplitudes (Fig.~S\ref{fig:magneticfield}b), we extract $|\Delta_g|=0.55 \pm 0.04$, which is in close agreement with the theoretical prediction \cite{MacQuarrie2021}.

\subsection{Thermal broadening}

\begin{figure}[h!]
    \centering
    \includegraphics{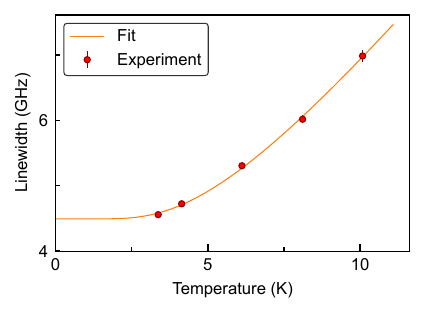}
    \caption{\textbf{Thermal broadening}. The measured temperature-dependent linewidth is fitted using Eq.~\ref{eq:thermalBroadening}. The fitted parameters are $E_a$=1.35 $\pm$ 0.02 meV, $P_\text{T}$=9.21 GHz, $P_0$=4.49 GHz.
    }
    \label{fig:thermal broadening}
\end{figure}

The cavity-coupled single T center emission line can also be thermally broadened due to activation of the TX$_0$-TX$_1$ transition \cite{Bergeron2020PRXQuantum}. To estimate the thermal broadening, we measure the temperature-dependent linewidth for the single T center (Fig.~S\ref{fig:thermal broadening}) and fit the results using, 

\begin{equation}
    \Gamma (T) = P_0 + \frac{P_\text{T}}{e^{Ea/k_BT}-1},
    \label{eq:thermalBroadening}
\end{equation}

\noindent where $P_0$ is the linewidth at $T=0$ K, $P_\text{T}$ is the thermal broadening coefficient, and $E_a$ is the activation energy. The fitted $E_a = 1.35 \pm 0.02$ meV is close to the reported 1.76 meV TX$_0$-TX$_1$ state splitting \cite{Bergeron2020PRXQuantum}. The comparison between $P_0$ and the measured linewidth of the cavity-coupled single T center at $T=3.4$ K suggests a thermal broadening value of $\Gamma_\text{th} \sim 0.1 $ GHz.


\subsection{Lifetime analysis}

\begin{figure}[h!]
	\centering \includegraphics{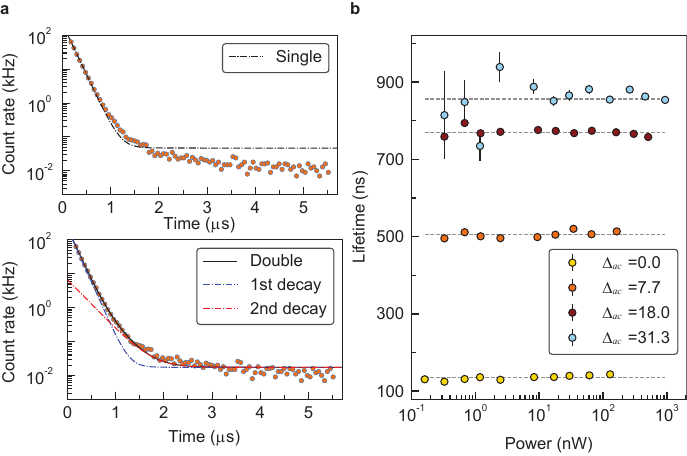}
    \caption{\textbf{Fluorescence decay lifetime analysis.} 
    \textbf{a.} Single- and bi-exponential fitting for the cavity-coupled single T center fluorescence decay at $\Delta_{ac} = 0$ (Fig.~4a of the main text). Single-exponential fitting (upper panel) reveals $\tau_1=144.3$ ns, with a count-rate amplitude $C_\text{max}^1=213.4$ kHz. Bi-exponential fitting (lower panel) reveals $\tau_1=136.4$~ns and $\tau_2=298.1$ ns, with count-rate amplitudes $C_\text{max}^1=217.1$ kHz and $C_\text{max}^2= 6.7$ kHz, respectively.
    \textbf{b.} Fluorescence lifetime of the cavity-coupled single T center under different excitation powers at certain detuning $\Delta_{ac}$.
    For the $\Delta_{ac}= 0$ case, we perform bi-exponential fitting and use the lifetime for the major decay. The average lifetime for each detuning is 134.7, 505.3, 770.8 and 856.5 ns. 
    }
	\label{fig:lifetimePower}
\end{figure}

Due to the spectral overlap between the cavity-coupled single T center with the inhomogeneous distribution of the waveguide-coupled T centers, background T center emission is expected while measuring the fluorescence decay of the cavity-coupled T center 
(Fig.~4a of the main text). Here we compare the fitting results of the decay data using single- and bi-exponential functions (Fig.~S\ref{fig:lifetimePower}a). The single-exponential fitting reveals a fluorescence decay lifetime $\tau_\text{exc}=144.3 \pm 0.2$ ns, yet showing a large deviation at longer time scales. With bi-exponential fitting, the dominant decay has a fitted lifetime of $\tau_\text{exc}=136.4 \pm 0.6$ ns with an amplitude weight of 97.0\%. 
For the cavity-coupled single T center, the measured lifetime has no dependence on the excitation power (Fig.~S\ref{fig:lifetimePower}b), which rules out any forms of stimulated emission process.


    
\section{Numerical modeling of the coupled system} \label{numericalmodeling}

\subsection{Lindblad master equation} \label{Lindbladian}

The dynamics of the coupled system for a single T center in a cavity can be well described by the Jaynes-Cummings Hamiltonian $H_\text{JC}$ of the form,

\begin{equation}
    H_\text{JC}/\hbar = H_0 + H_d,
    \label{eq:JC_ham}
\end{equation}
\begin{equation}
    H_0=\Delta_a\sigma_+\sigma_- 
    + \Delta_c a^\dagger a 
    + g(\sigma_+ a + \sigma_- a^\dagger), H_d=\frac{\Omega}{2}(\sigma_+ + \sigma_-),
    \label{eq:JC_ham0}
\end{equation}

\noindent which assumes rotating wave approximation and is in the rotating frame of the laser field ($\omega_L$). Here $\Delta_a = \omega_a - \omega_L$ and $\Delta_\text{c} = \omega_c - \omega_L$ are, respectively, the detunings of the laser from the T center transition $\omega_a$ and from the cavity resonance $\omega_c$, $g$ is the coupling strength between the single T center and the single cavity mode, $\Omega$ is the optical Rabi frequency, $a$ and $\sigma_-$ are the annihilation operators for the cavity mode and the T center optical transition, respectively. The optical Rabi frequency $\Omega = 2g\sqrt{N_\text{ph}}$, where $2g$ is the single photon Rabi frequency and $N_\text{ph}$ is the average intracavity photon number. The $N_\text{ph}$ is related to the input power $P_\text{in}$ by the following equation  \cite{tiecke2014nanophotonic},

\begin{equation}
    N_\text{ph}=4\frac{\eta_\text{cav}/\kappa}{1+(2\Delta_c/\kappa)^2}\frac{P_\text{in}}{\hbar\omega_a},
    \label{eq:intraCavityPhoton}
\end{equation}

\noindent where $\kappa$ is the cavity linewidth. We note the equivalence between semiclassical drive $H_d=\frac{\Omega}{2}(\sigma_+ + \sigma_-)$ and coherent drive $H_d=\sqrt{\frac{\kappa_\text{wg}P_\text{in}}{\hbar\omega_L}} (a^{\dagger} + a)$, given a sufficiently large photon Fock state space. To analyze the open quantum system, we utilize the Lindblad master equation in QuTiP \cite{johansson2012qutip} to describe the time evolution of the light-matter interaction,


\begin{equation}
    \frac{d}{dt}\rho = \frac{-i}{\hbar}[H_\text{JC}(t),\rho]
    +\sum_{i} C_i\rho C_i^\dagger 
    -\frac{1}{2}\{C_i^\dagger C_i, \rho\} \equiv \mathcal{L}\rho.
    \label{eq:lindbladian}
\end{equation}

The jump operators $C_i$ describe different loss mechanisms. We define major loss channels as $\sqrt{\kappa}a$, $\sqrt{\Gamma_0}\sigma_- $, and $\sqrt{\Gamma_\text{d}/2}\sigma_z$, which respectively represent cavity decay, T center spontaneous emission, and pure dephasing with a rate of $\Gamma_\text{d}$. To simulate the pulsed excitation, we implement a time-dependent Hamiltonian $H(t)$, which consists of $H_0+H_d$ during the excitation and $H_0$ when the excitation is off. To take into account the possible spectral diffusion ($\Gamma_\text{sd}$) of the single T center, we perform a weighted average of the simulation outcomes assuming $\Delta_a$ takes a range of values with each weight determined by a Gaussian distribution with a FWHM linewdith of 2$\Gamma_\text{sd}$.

\begin{figure}[h!]
	\centering
    \includegraphics{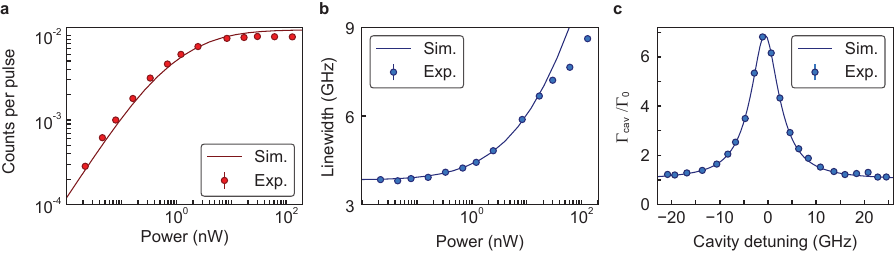}
    \caption{\textbf{Lindbladian calculation results.} Global numerical fitting of the saturation, linewidth and detuning-dependent fluorescence decay data (reproduced from Fig.~3c, Fig.~3d, and Fig.~4b of the main text) by solving the Lindblad master equation, using $g$, $\Gamma_\text{d}$ and $\Gamma_\text{sd}$ as the fitting parameters.
    }
	\label{fig:Lindbladian}
\end{figure}

We extract the $g$, $\Gamma_\text{d}$ and $\Gamma_\text{sd}$ by fitting counts saturation, power-dependent linewidth, and detuning-dependent fluorescence decay data globally using the SciPy built-in function Basin-hopping and Nelder-Mead method \cite{virtanen2020scipy}, which reveals ($g$, $\Gamma_\text{d}$, $\Gamma_\text{sd}$) =$2\pi\times$(42.4 MHz, 0.65 GHz, 1.69 GHz). 
The global fitting can match with the different measurement data simultaneously (Fig.~S\ref{fig:Lindbladian}). The characteristic linewdith of $\tilde{\kappa}$ has contributions from the cavity linewidth $\kappa$, the dephasing $\Gamma_\text{d}$, and the spectral diffusion $\Gamma_\text{sd}$. In a hypothetical scenario where $\Gamma_\text{sd}$ is negligible, we can have an analytical formula for $\tilde{\kappa} \approx \kappa + 2\Gamma_\text{d}$ derived from the numerical model mentioned above.

\subsection{Cavity-coupled spectrum}

\begin{figure}[h!]
	\centering
    \includegraphics{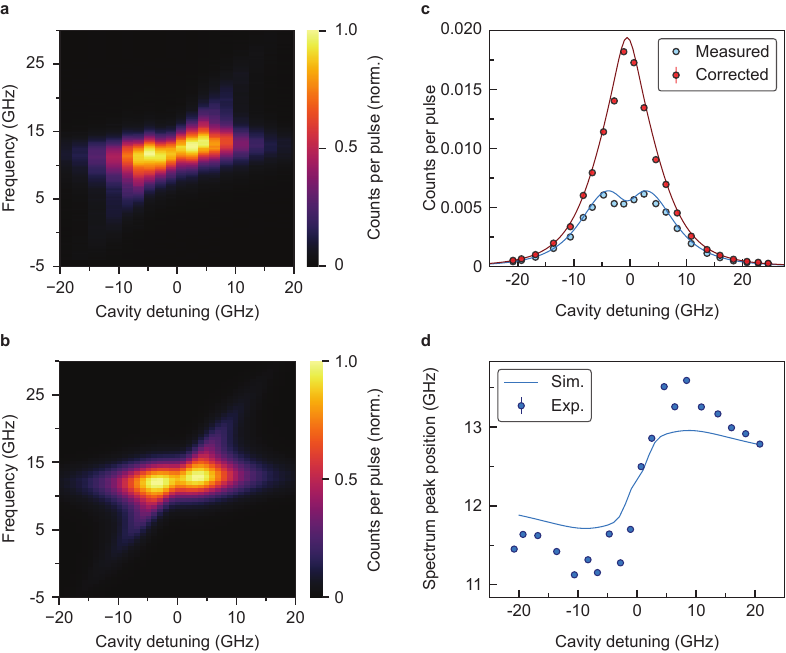}
    \caption{\textbf{Cavity-coupled single T center spectrum.}
    \textbf{a.} Measured 2D spectrum map for the cavity-coupled single T center at different $\Delta_{ac}$ under an excitation power $P_\text{in} = 1.21$ nW.
    \textbf{b.} Simulated 2D spectrum map based on the numerical calculations.
    \textbf{c.} Measured and corrected counts at the T center optical transition ($\omega_a$) at different $\Delta_{ac}$. The correction factor is $e^{t_0/\tau_\text{exc}}$. The solid lines are the results from the numerical calculations.
    \textbf{d.} Extracted PLE spectrum peak positions at different $\Delta_{ac}$. The blue line represents the numerical calculation results.
    }
	\label{fig:spectrum_cavitydetuning}
\end{figure}

As the cavity resonance is tuned across the cavity-coupled single T center optical transition, we observe modulation of the T center spectrum depending on the cavity detuning $\Delta_{ac} = \omega_c - \omega_a$ (Fig.~S\ref{fig:spectrum_cavitydetuning}a). This is mainly attributed to the dependence of the optical excitation on the $\Delta_c$, which results from the comparable linewidth of the single T center and the cavity ($\Gamma \sim \kappa$). The driving strength across the T center linewidth varies because the average intracavity photon number $N_\text{ph}$ depends on the $\omega_L$, which can be seen by rewriting Eq. \ref{eq:intraCavityPhoton} as,

\begin{equation}
    N_\text{ph}= 4\frac{\eta_\text{cav}/\kappa}{1+[2(\Delta_{ac}+\omega_a-\omega_L)/\kappa]^2}\frac{P_\text{in}}{\hbar\omega_L}.
    \label{eq:intraCavityPhoton_rewrite}
\end{equation}

\noindent For a specific cavity detuning $\Delta_{ac}$, when $\omega_L$ is scanned (with a typical range $\sim 5\Gamma$) to obtain the single T center PLE spectrum, $N_\text{ph}$ will be modulated strongly. In a hypothetical regime where $\Gamma \ll \kappa$, this effect will disappear.

We turn to the numerical calculations to verify the observed 2D spectrum map (Fig.~S\ref{fig:spectrum_cavitydetuning}a). We use the fitted ($g$, $\Gamma_\text{d}$, $\Gamma_\text{sd}$) determined in Section \ref{Lindbladian} for solving the Lindblad master equation to predict the single T center emission spectrum at different cavity detunings. The predicted 2D spectrum map (Fig.~S\ref{fig:spectrum_cavitydetuning}b) match well with the measurements. The lower count level at $\Delta_{ac}=0$ is due to the $t_0 = 170$ ns time window where the fluorescence cannot be collected (see Section \ref{collectefficiency}), which affects more at zero detuning as the T center has a shorter fluorescence lifetime. The numerical calculations also well capture the raw and corrected count level (Fig.~S\ref{fig:spectrum_cavitydetuning}c) at the T center optical transition ($\omega_a = 12.34$ GHz), as well as the evolution of the peak position for the cavity-coupled PLE spectrum at different cavity detunings (Fig.~S\ref{fig:spectrum_cavitydetuning}d).





\clearpage
\setlength\bibitemsep{0pt}
\bibliography{SingleT_SI_references}